\documentclass{aastex}
\begin{document}
\newcommand\kms{~km~s$^{-1}$}
\shorttitle{Progenitor Masses of Wolf-Rayet Stars}
\shortauthors{Massey, DeGioia-Eastwood, and Waterhouse}
\title{The Progenitor Masses
of Wolf-Rayet Stars and
Luminous Blue Variables
Determined from Cluster Turn-offs. II. Results from 12 Galactic Clusters
and OB Associations}

\author{Philip Massey\altaffilmark{1,}\altaffilmark{2}}
\affil{Kitt Peak National Observatory, National Optical Astronomy
Observatory\altaffilmark{3}}
\affil{P.O. Box 26732, Tucson, AZ 85726-6732}
\email{massey@lowell.edu}

\author{Kathleen DeGioia-Eastwood\altaffilmark{4}
and Elizabeth Waterhouse\altaffilmark{4,}\altaffilmark{5}}
\affil{Department of Physics and Astronomy, Northern Arizona University
\\P.O. Box 6010, Flagstaff, AZ 86011-6010.}
\email{kathy.eastwood@nau.edu}
\email{waterh@fas.harvard.edu}

\altaffiltext{1}{Present address: Lowell Observatory, 1400 W. Mars Hill
Road, Flagstaff, AZ 86001.}
\altaffiltext{2}{Visiting Astronomer, Cerro Tololo Inter-American
Observatory, National Optical Astronomy Observatory (NOAO), which is operated by
the Association of Universities for Research in Astronomy, Inc. (AURA) under
cooperative agreement with the National Science Foundation (NSF).}
\altaffiltext{3}{Operated by AURA
under cooperative agreement with the
NSF.}
\altaffiltext{4}{Visiting Astronomer, Kitt Peak National Observatory, NOAO,
which is operated by AURA under cooperative agreement with the NSF.}
\altaffiltext{5}{Participant in the NSF's Research Experiences for 
Undergraduates program,
Summer 1998. Present address:
Harvard University, 436 Eliot House Mail Center, 101 Dunster Street, Cambridge, MA 02138.}

\begin{abstract}

In a previous paper on the Magellanic Clouds we
demonstrated that coeval clusters provide a powerful
tool for probing the progenitor masses of Wolf-Rayet stars (W-Rs)
and Luminous
Blue Variables (LBVs).  Here we extend this work to the higher metallicity
regions of the Milky Way, studying 12 Galactic clusters. We present new
spectral types for the unevolved stars and use these, plus
data from the literature, to construct H-R diagrams.  We find that all
but two of the clusters are highly coeval, with the highest mass stars
having formed over a period of less than 1~Myr.  The turn-off masses
show that at Milky Way metallicities some W-Rs (of early WN type)
come from
stars with masses as low as 20--25$\cal M_\odot$. Other early-type WNs
appears to have evolved from high masses, suggesting a large range of masses
evolve through an early-WN stage.
On the other hand,
WN7 stars are
found only in clusters with very high turn-off masses, $>120\cal M_\odot$.
Similarly the LBVs are only found in clusters with the highest turn-off masses,
as we found in the Magellanic Clouds, providing very strong evidence that
LBVs are a normal stage in the evolution of the most massive stars.
Although clusters containing WN7s and LBVs can be as young as
1~Myr, we argue that these objects are evolved, and that the
young age simply reflects the very high masses that characterize the
progenitors of such stars. In particular we show that the 
LBV $\eta$~Car appears to be coeval
with the rest of the Trumpler~14/16 complex.
Although the WCs in the Magellanic Clouds were found in clusers with
turn-off masses as low as $45\cal M_\odot$, the three Galactic WCs
in our sample are all found in clusters with high turn-off masses
($>70\cal M_\odot$); whether this difference is 
significant or due to small-number statistics remains to be seen. The BCs
of Galactic W-Rs are hard to establish using the cluster turn-off method,
but are consistent with the ``standard model" of Hillier. 

\end{abstract}


\section{Introduction}
\subsection{Background}

Massive, luminous stars begin their H-burning lives as hot, O-type stars.
During their main-sequence evolution (2.5--8~Myr for stars with initial
masses of 120--20$\cal M_\odot$) 
they may lose a significant
amount of their mass due to strong stellar winds.  The observed mass-loss
rates suggest that the highest-mass stars will lose as much as half their
mass during the H-burning stage. Since these winds are driven by radiation
pressure through highly-ionized metal lines, the mass-loss rates will 
increase with stellar luminosity, metallicity, and effective temperature.
It was Conti (1976) who first suggested that
mass-loss provided a simple explanation for how Wolf-Rayet (W-R) stars form.
In the modern version of the ``Conti scenario" (Maeder \& Conti 1994),
this mass loss results in a stripping off of the H-rich outer layers
of the star, resulting in a WN-type W-R star,
in which the H-burning products (He and N) are enriched at the 
surface, with strong, broad emission lines indicative of enhanced stellar
winds forming 
an extended atmosphere.  Most WNs are presumed to be He-burning objects,
although there is evidence that a few H-rich late-type WNs 
are still near the end of
the core-H burning phase (Conti et al.\ 1995).
Further mass-loss and evolution reveals 
the products of He-burning (C and O) at the surface, and the star is
spectroscopically identified as a WC-type W-R. 

Possibly the
highest mass stars also go through a Luminous Blue Variable (LBV) stage
on their way to becoming W-Rs, with the large, episodic mass loss that
characterizes LBVs aiding the process. 
Stars of slightly lower luminosity (mass) will not have an LBV phase,
but recent observations (Massey 1998a) suggest that they do go through an
intermediate red
supergiant (RSG) phase, even at high metallicities, albeit for a short
time. At some
lower 
luminosity one expects that the mass-loss rates are sufficient to produce a WN-type
W-R, but not a WC.  And, at even lower luminosities, mass-loss rates are so
low that the He-burning stage
is spent entirely as a red supergiant (RSG) and not as a W-R star.  

In this
version of the ``Conti scenario" we thus might
expect the following ``paths" to be
followed, in order of decreasing luminosities:

\begin{centering}

O$\longrightarrow$LBV$\longrightarrow$WN$\longrightarrow$WC (1)

O$\longrightarrow$RSG$\longrightarrow$WN$\longrightarrow$WC (2)

O$\longrightarrow$RSG$\longrightarrow$WN (3)

OB$\longrightarrow$RSG (4)

\end{centering}

The masses corresponding to these paths are unknown (and indeed we are
unsure even of the qualitative correctness of these paths), but ``standard
guesses" for characteristic values would be 1: $\geq80\cal M_\odot$, 2: $60\cal M_\odot$, 3: $40\cal M_\odot$, and 4: $20\cal M_\odot$.  We emphasize that these are purely
speculative values, and that the actual ranges should depend upon metallicity.
Indeed it is to address this issue that the present series of papers has
come about.

Between the initial O-type stage (of luminosity class ``V")
and the He-burning stage (LBVs, W-Rs, RSGs) the star will become a
B-type supergiant; most, but not all, of these are expected to still
be H-burning objects (Massey et al.\ 1995b).  The luminous A-F supergiants
are very short-lived intermediate stages during He-burning for stars
of intermediate high-mass, depending upon
the metallicity.
The precursor to SN1987A is
believed to have been a ``second-generation" B-type supergiant, a He-burning
object of somewhat lower mass than those being discussed here.

In addition to the question of whether or not the above paths are correct,
and what masses to assign to each as a function of metallicity, we are also
interested in the evolutionary significance, if any, to the Wolf-Rayet
spectral subtypes. WN-type W-R stars are
classified as ``early" (WNE) or ``late" (WNL) depending on whether NV~$\lambda\lambda 4602,19$ or
NIII~$\lambda\lambda 4634,42$ dominates; e.g., 
WNEs correspond to numerical subtypes WN2, WN3, WN4, while
WNLs consist of subtypes WN6, WN7, WN8, and WN9, with the
WN5 stars split between the two groups. Similarly WC-type
W-R stars are classified as WCE or WCL depending upon whether CIII~$\lambda 5696$
dominates over CIV~$\lambda 5808$; e.g., corresponding to spectral subtypes
WC4 through WC6 and WC8-9, with some WC7s falling into each camp 
(Conti \& Massey 1989).  Do these subtypes mean anything
in an evolutionary sense?  Various authors have claimed so (see,
for example, Moffat 1982), but this conjecture
does not seem borne out by either observational or theoretical studies. 

Our understanding of massive star evolution is limited, in part, because of the
difficulty of constructing stellar models from first principles.  The physics
of massive stars is complicated by strong stellar winds, and the choice of
the functional dependence of mass-loss rates on stellar parameters
(luminosity, temperature, mass, and metallicity)
greatly influences the
theoretical tracks (e.g., Meynet et al.\ 1994), particularly in the later
stages of evolution.  In addition, the models are sensitive to the amount
of mixing. However, there is little agreement
on the treatment of the relevant processes of semi-convection and overshooting
(Maeder \& Conti 1994), while the most recent work has emphasized the
significant role that rotation may play in this regard (Maeder 1997, 1999).
Nevertheless, empirical studies of massive star evolution provide confidence
that the above picture is correct, and are beginning to provide quantitative
information on the mass ranges corresponding the various paths.  These
studies provide an observational basis against which the models can be
evaluated and refined. (For a humorous rendering of this process, the
reader is referred to Fig.~5 in Conti 1982.)

\subsection{An Observational Approach}
\subsubsection{Global Studies}

The galaxies of the Local Group provide perfect laboratories for pursing these
studies observationally, as the metallicity differs by almost an order of
magnitude (SMC to M31) among the galaxies currently active in forming stars.
During the past few years there have been a number of studies of 
{\it mixed-age} populations, the relative number of {\it this} and {\it that}.
The implicit assumption of these studies are that the IMF
slopes are identical and cover regions that provide a good sampling
of stellar stages over time.  The number ratios provide quantitative criterion
for the models to attempt to match. 
These studies have found the following:

(1) The relative number of RSGs to W-Rs decreases with increasing metallicity
(Massey 1998a; Massey \& Johnson 1998).
Histograms of the number of RSGs vs.\ luminosity reveal that there
are proportionately fewer high luminosity RSGs at higher metallicity,
while the lack of a sharp luminosity cut-off
supports the interpretation that as $Z$ increases,
massive stars spend a greater fraction of their He-burning phase as
W-Rs, and a smaller fraction as RSGs, rather than there being a difference
in the actual mass ranges that go through a RSG phase.  This is why we indicated
a RSG phase at high luminosities (path 2 above). 
Possibly even LBVs will go---or have gone---through a RSG phase, but this
is unknown.
We also do not know if massive stars go through a RSG phase
at the highest metallicity: the relation between the
RSG/W-R ratio and metallicity appears to flatten below the high metallicity
 that characterize M31, but only a few regions have
been surveyed in that large galaxy, and more data are being
gathered to resolve this.

(2) The relative number of WCs to WNs increases with increasing metallicity,
with the notable exception of the star-burst galaxy IC~10 (Fig.~8 in
Massey \& Johnson 1998).  This trend is
also
in accord with the predictions of the Conti scenario, as increased mass-loss
makes it possible for a star of a given luminosity to reach the WC stage
sooner, spending more of its W-R stage as a WC than a WN. (The explanation for IC10's peculiarly high WC/WN ratio remains a mystery at present. See 
discussion in Massey \& Johnson 1998.)

(3) Significant differences in the spectral subtypes found in the Magellanic
Clouds compared to the Milky Way were noted by Smith (1968): no WCLs are
found in the LMC or SMC, and most of the ones known in the Milky Way are
found inwards of the solar circle. Similar differences are seen for the WNs.
Armandroff \& Massey (1991) showed that the WC line widths (which are correlated
with spectral subclass) change systematically with metallicity, extending an
important finding by Willis, Schild, \& Smith (1992) to other galaxies of
the Local Group. Massey (1996) proposed that the WC spectral subtypes
are nothing more than an atmospheric effect due to metallicity. (See also
Massey \& Johnson 1998).  Recently Crowther (2000) has demonstrated from
W-R model atmospheres that the WN subtypes may similarly be a reflection of
metallicity rather than other stellar parameters, at least in terms of WN3
through WN6.

\subsubsection{Coeval Associations}
A more direct way exists to attack the problem of understanding
massive star evolution.  By using coeval associations that contain
evolved, massive stars, we can in fact directly 
{\it measure} the mass ranges that correspond
to the above evolutionary paths as a function of metallicity. This
is the subject of the current series of papers.

In Paper I of this series (Massey, Waterhouse, \& DeGioia-Eastwood 2000) we established that many Magellanic Cloud OB associations and young
clusters are sufficiently coeval ($\Delta \tau<$1~Myr) that we can measure
a meaningful cluster turn-off mass; e.g., the mass of the most massive
star on the main-sequence.  These turn-off masses then place
a lower-limit on the mass of the progenitors of the
evolved stars in the cluster, to the extent that star-formation proceeded
coevally. If the cluster is well populated, than the initial mass of the
turn-off star is a good approximation to the initial mass of the
progenitor of the W-R star.
In addition, the bolometric luminosity of the cluster
turn-off sets useful limits on the bolometric corrections
for the evolved stars, allowing tests of Wolf-Rayet model atmospheres,
such as Hillier's ``standard model" (Hillier 1987, 1990).

The results were quite revealing. In a study of 19 OB
associations in the Magellanic Clouds, we found that at the
low metallicities
that characterize the SMC, only the highest-mass stars ($>70 \cal M_\odot$) become Wolf-Rayet stars, although the sample is  small.
This is equivalent to saying that path (3) above occurs only for $M>70\cal M_\odot$ for $Z\leq 0.005$.\footnote{For convenience in talking about the metallicity $Z$, we
adopt $Z=0.018$ for the solar neighborhood, corresponding to log~O/H+12=8.70
(Estebam \& Peimbert 1995).  If we then simply scale $Z$ relative to the
easily-measured oxygen abundance, $Z=0.005$ for the SMC (log~O/H+12=8.13) and $Z=0.008$ for
the LMC (log~O/H=8.37), e.g., Russell \& Dopita (1990).  Although it is well
recognized that different metals will have different relative abundances, it
is fortuitous that it is oxygen (along with carbon and nitrogen) which
are the primary accelerators of the stellar winds at the high effective
temperatures appropriate to O-type stars (Abbott 1982).}
At the higher metallicity of the LMC ($Z=0.008$), WN W-R stars come from stars with masses as low as $30\cal M_\odot$.
We also found that WC stars are found
in the same clusters as WNEs; e.g., the lowest turn-off mass found for
a cluster containing WC stars was 45$\cal M_\odot$, suggesting that stars
with masses from 30--45$\cal M_\odot$ might correspond to path (3), 
while stars with masses 45-85$\cal M_\odot$ correspond to path (2).

The rare ``Ofpe/WN9" stars\footnote{The stars may be
rarer at higher metallicity than at low; ten
are known in the LMC (Bohannan \& Walborn 1989), while 
only one is known in the Milky Way (Bohannan \& Crowther 1999). Six
are known in M~33 (Massey et al.\ 1996), while one is known in the higher
metallicity M~31 (Massey 1998b).},
once thought to be a transition type between ``Of" stars and Wolf-Rayet stars,
are only found in clusters and associations with the lowest turn-off
masses, 25-35$\cal M_\odot$.  Recently the Ofpe/WN9 stars were implicated
in the LBV phase of massive stars, after one Ofpe/WN9 star
(R127) underwent an ``LBV-like" outburst.  But, the classical LBVs in our
LMC sample, including the archetype itself S Doradus, are found in clusters
with the very highest turn-off masses, $>85 \cal M_\odot$.  (Similarly the
SMC W-R star HD~5980, which many consider to be a ``true" LBV [Barb\'a et al.\ 1995] is found in a cluster with a very high turn-off mass.) We conclude
that the
Ofpe/WN9 stars are just the lowest-mass versions of Wolf-Rayet stars.
True LBVs, on the other hand, are found only in the clusters with the 
highest mass turn-offs, suggesting that they are indeed stars near their
Eddington limit and are a normal stage of the most massive stars.

Our study also shed light on the origin of the different W-R classes
and subtypes, at least at the modest metallicity
that characterizes the LMC. The early-type WN stars (WNEs) in the LMC are found
in clusters with a large range of turn-off masses (from 30$\cal M_\odot$ to
100$\cal M_\odot$ or more), suggesting that these are a stage that most massive
stars go through at LMC-like metallicities.

We turn now to the higher metallicity of our own Milky Way, and pose the
question of where do LBVs and Wolf-Rayets of various types come from at
a metallicity considerably higher than that of the Magellanic Clouds.

\section{The Sample}

Previous attempts to use Galactic OB associations and clusters to measure the
progenitor masses of W-R stars have been made by Schild \& Maeder (1984), 
Humphreys, Nichols \& Massey (1985), 
and Vazquez \& Feinstein (1990); Smith, Meynet, \& Mermilliod (1994) also 
discuss the issue but primarily with an emphasis on using the
data for bolometric corrections.  These studies relied upon results
from the literature and did not obtain new spectroscopy of the cluster stars. Our experience, even in nearby young clusters, is that many
of the high mass stars have been overlooked.\footnote{For example, 
Massey \& Thompson (1991) identified numerous O-type stars previously missed in 
Cyg~OB2, including one as early as O4~III(f). Similarly spectroscopy by
Hillenbrand et al.\ (1993) of NGC~6611 found stars that had been
previously called ``O4" were in fact of ``B0~V" type (NGC~6611-188) in one
case and ``O5~V" in another (NGC~6611-205=HD168076), and provided modern
spectral types for dozens of others. Massey \& Johnson (1993) found another
O3~If (Tr14/16-506) even in the well-studied $\eta$ Carinae region, and provided
new spectral types for others.  Table~3 of Massey, Johnson, \& Eastwood
(1995a) gives numerous other examples of newly discovered early-type O stars
in nearby Galactic regions.
Spectroscopy is of course critical for assigning location of hot stars
accurately in the HRD, as emphasized in Paper~I and elsewhere (e.g., 
Massey et al.\ 1995a, 1995b).}

In this paper we draw upon the literature, but also obtain
new spectroscopy for the regions where this is required.  We note that few of
these regions have modern photometry.  This is largely irrelevant for our
purposes, as discussed further in Section~\ref{Sec-dist} below, but does
affect any attempts to use the distances as probes of Galactic structure.

\subsection{Selection}

For our sample, we began with the lists of W-Rs believed to be likely members
of clusters and associations given by Lundstrom \& Stenholm (1984) and
Schild \& Maeder (1984).  We eliminated the many regions that lacked 
{\it UBV} photometry (unfortunately this excludes many fine southern
regions), were too big (NGC~2439, Vul~OB2), or were 
too sparse (the ``HD~155603
Group").  This left us with 12 associations for which we either obtained
new spectroscopy, or adequate data existed in the recent literature.
We list these regions in Table~1.

Two regions not listed in the table require  
special comment.  First, we have excluded the region
NGC~3603 from our study.  NGC~3603 is the Milky Way's answer to R136, in
that this is a young (1-2~Myr), rich region that is so highly populated
that the IMF extends up to very high masses. In both cases {\it HST} was
needed for spatially resolving a single ``WN+abs" star into multiple
objects, and for obtaining spectra of the individual components.
NGC~3603 
contains several stars
with WN6-like features, but whose individual luminosities are much higher
than that of normal W-Rs of their type, and whose spectra show unmistakable
evidence of hydrogen, also not in accord with their type. (This can
be inferred from Fig.~2 in  Drissen et al.\ 1995, although the significance
at the the time was not apparent.) 
Massey \& Hunter (1998) found the same thing in R136,
and realized that these WN stars were
simply ``super Of" stars---core-H burning objects whose very large 
luminosities (corresponding to masses well above the 120$\cal M_\odot$ limits
of published evolutionary tracks) result in such strong stellar winds that
their spectral appearance mimics that of Wolf-Rayet stars. (The same conclusion
is reached  by de Koter, Heap, \& Hubeny 1997.)
An excellent comparison between the NGC~3603 and R~136 objects 
is provided by Crowther \& Dessart (1998). We
ignore these objects here, primarily as we do not consider them true Wolf-Rayet stars.

We also call attention to the
cluster NGC~6231 (the nucleus of the 
large Sco OB1 association), which contains two Wolf-Rayet stars, the WN7 star
HD151932, and the WC7+O5-8 binary HD155720.  Although the cluster has
received recent photometric attention (Perry, Hill, \& Christodoulou 1991;
Sung, Bessell, \& Lee 1998;
Baume, Vazquez \& Feinstein 1999), there is no
 modern {\it spectroscopic} study of this 
highly interesting region.  Such a study would bolster or refute claims
of large age spreads and peculiar mass functions in this cluster, as well as 
potentially providing 
additional information on the progenitor masses of Wolf-Rayet
stars.

\subsection{New Data: Spectral Types and Improved Distances and Reddenings}

\subsubsection{Spectral Types: New Data and Classification}

Owing to the recent study of a number of northern hemisphere OB young clusters
and OB associations (Massey et al.\ 1995a and references therein), most of our
new spectroscopy was obtained for stars in interesting southern OB
associations. The majority of the new spectra were obtained on the CTIO
1.5-m telescope with the Loral 1K CCD (1200 $\times$ 800 pixels)
spectrometer on 1998 Mar 19 (UT) with grating 47 in second order
and a CuSO$_4$ blocking filter.  The dispersion was 0.56~\AA~pixel$^{-1}$,
and the 100~$\mu$m slit (1.8 arcsec) yielded a resolution of 
1.5~\AA\ (2.5 pixels) with coverage from
4035~\AA\ to 4700~\AA.

Data were also obtained from Kitt Peak 
telescopes of the northern clusters (Berkeley~87 and Ma~50), as well as
critical data for some stars in the southern associations,
admittedly at very low elevations.  Most of these were obtained on the
Mayall 4-m during 1998 Sept 11-14. The RC Spectrograph was used with
the T2KB detector (2048$\times$ 2048 pixels), with grating KP-22 used
in second order and a CuSO$_4$ blocking filter. The dispersion was
0.72~\AA\ pixel$^{-1}$, and the 300~$\mu$m (2.0 arcsec) 
slit yielded a resolution of 1.8~\AA\ (2.5 pixels), with coverage from
3750~\AA\ to 5000~\AA\ being in good focus.

A few data were also obtained on the Kitt Peak 2.1-m telescope
1998 July 19 and 21, and on 2000 Mar 17, using the ``GoldCam" spectrometer
with its Loral 3K x 1K  CCD.  Grating 47 was used in second order with
a dispersion of 0.47~\AA~pixel$^{-1}$, and the 250~$\mu$m (3 arcsec)
slit provided a resolution of 1.9~\AA\ (4 pixels).
The wavelength coverage was 3800~\AA\ to 4800~\AA.

Thus in all cases the wavelength range covered the important classification
lines Si~IV~$\lambda 4089$ to He~II~$\lambda 4686$, at resolutions of 
1.5--1.9~\AA.

Good
flat-fielding was provided by exposures of an illuminated white spot.
Wavelength calibration was by means of He-Ar (CTIO) or He-Ne-Ar (KPNO)
lamps.  The customary IRAF optimal-extraction routines were used.
We usually achieved a S/N of 80 per spectral resolution element.
We classified the stars based upon 
the criterion given in Walborn \& Fitzpatrick (1990).

We list in Table~2 the brightest stars in each of these associations,
including our new spectral types
plus any from the literature.  We have measured accurate coordinates from
the Space Telescope Science Institute's Digitized Sky Survey images, and
include these in Table~2 to facilitate exact identification.  We describe
the most interesting spectra here.

\paragraph{Two Newly Discovered O3 Stars.}
\label{Sec-o3s}  The ``O3" class was introduced by
Walborn (1971) as an extension of the MK classes to hotter effective
temperatures. At modest resolution and signal-to-noise (S/N), the spectra show
no He~I lines, but rather strong He~II.  The class is clearly degenerate,
as higher resolution and better
S/N shows He~I $\lambda 4471$ with equivalent widths as
large as 120-250~m\AA\ in some stars (Kudritizki 1980; Simon et al.\ 1983), and less than 75~m\AA\ in others (i.e., Paper~I).  Since the O3 class represents the
hottest class, all members are of high mass, and stars of type O3~III and
O3~I must be of extremely high mass.  Such stars are correspondingly rare,
with only five possible examples in the Milky Way (e.g., the four mentioned
by Walborn 1994 plus one other possible example discovered by
Massey \& Johnson
1993 in Tr~14/16).  

Thus,
our discovery here of two additional such stars, both in the poorly studied
cluster Pis~24 (Sec.~\ref{Sec-Pis24}) is of some interest.  We illustrate
the spectra of these two stars, Pis24-1 and Pis24-17 in Fig.~\ref{fig:o3s}.  
In the case of
HDE~319718=Pis24-1, we do detect He~I $\lambda 4471$ with an equivalent width
of 85~m\AA, comparable to that seen in the most extreme of Carina stars which
originally defined the class.
Pis24-1 is clearly of type O3~If*: the ``f" as both N~III $\lambda \lambda 4634, 42$ and
He~II $\lambda 4686$ are seen in emission, and the ``*" signifying N~IV~$\lambda 4058$ emission stronger than N~III. 
Both features are luminosity indicators, thus resulting
in the ``I" luminosity class.  Note also the strong N~V $\lambda 
\lambda 4603, 19$
absorption, which also appears to be stronger in O3 stars of high luminosity,
although one might expect there is also a strong temperature dependence for
this line, which is not seen in O4 stars, except at high luminosities.
The star had been
previously classified as ``O4(f)" by Lorret, Testor, \& Niemela (1984), although
in fact Vijapurkar \& Drilling (1993) had suggested an ``O3~III" designation.

In the case of Pis24-17, we do not detect He~I. We would argue that its
luminosity
class is intermediate between ``III" and ``I": on the one
hand, 
He~II $\lambda 4686$ is strongly in absorption while N III is in emission,
suggesting an O3~III(f*) classification, given that the N~IV emission is
comparable in strength to that of N~III. 
However, there is  incredibly strong N~V absorption, 
which would argue either for
higher luminosity (or higher effective temperature?).  Rather than attempt
to introduce a II(f*) into the nomenclature, we classify the star as O3~III(f*).
We are indebted to Nolan Walborn for his insightful comments on this spectrum.
This star was labeled "N35B'' by Lortet et al.\ (1984), and
classified as O4-5~V.  Doubtless the strong nebular He~I $\lambda 4471$
emission disguised the O3 nature of this and Pis24-1 on their photographic
spectra.

\paragraph{Other Early O-type Supergiants.}
We show in Fig.~\ref{fig:oIs} a few of our other early O-type supergiants.
The spectra here can be compared to those illustrated in Walborn \& Fitzpatrick (1990).
These spectra, and the new types listed in Table~2, emphasize the
need for modern spectroscopic studies of early-type stars even
within modest distances of the Sun.

\subsubsection{Reddenings and Distances}
\label{Sec-dist}

In order to locate the stars in the H-R diagram and assign masses, we need
to know their luminosities; for this,
we need to 
know their distances and  reddenings.   

We have photometry (from the literature), and good spectral
types (mostly new), 
for only the brightest dozen or so stars in each cluster (e.g.,
Table~2), with the exception of the clusters previously studied in order
to determine their initial mass function.  Main-sequence fitting is of little use when dealing with 
stars this hot, as color information (even in the absence of reddening)
provides little information about effective temperature.  Instead, we
derive distances and reddenings through the spectral types, adopting the
absolute magnitude calibration discussed in Paper~I, as well as the
same intrinsic color calibration.

For each star with a spectral type in Table~2, we computed
the reddening and true distance modulus, and then eliminated any obvious
outliers. Discrepant distances can be due
either to misclassification by a luminosity class (relatively
easy for the early B stars) or non-membership; multiplicity must also
play an occasional role. The results are given in
Table~1, along with the other data on the clusters.  We have also included
an estimate of the size of the region.
In general, the determination of the distance moduli is good to 0.1~mag\footnote{It is worth noting that our method of deriving the distances and reddenings
makes our results completely independent of any  zero-point errors in the photometry, 
either in $V$ or even in $B-V$.  For instance, imagine that the published
photometry was in error by 0.1~mag in $V$ ($V_{\rm true}=V_{\rm published}+0.1$) and by +0.1~mag in $B-V$ ($(B-V)_{\rm true}=(B-V)_{\rm published}+0.1$).  The
extinction correction $A_V=3.2\times E(B-V)$ will be underestimated by
0.32~mag, and the $V_o$ we compute will be too large by 0.22~mag, and our
value for the distance modulus too small by 0.22~mag.  However, when we go
to use this incorrect distance modulus with the rest of the photometry, for
which we do not have spectral types, it exactly compensates for the systematic
photometric errors, reproducing the correct values for $M_V$ and $(B-V)_o$.
This is of course because we determined the distance modulus and reddening
that caused us to match the correct (adopted) $M_V$ and $(B-V)_o$ based
upon the spectral type.  In a {\it reductio ad absurdum} the $V$ magnitudes
could be off by 10~mag, and the $B-V$ values off by 1~mag, and although our
distance moduli would be ridiculous, we would nevertheless be able to construct
accurate H-R diagrams, as long as the {\it relative} $V$ magnitudes and
{\it relative} $B-V$ values, are correct. It is easy to extend this argument
to $U-B$ and the color-free index $q_r=E(U-B)/E(B-V)$, that we will
in fact use.  In practice, errors in the published photometry probably include
color terms as well.  Redoing the photometry for the clusters using modern
CCD techniques would be well worthwhile, but primarily for improving our
knowledge of Galactic structure---refinements in the distance moduli and 
actual reddenings will have little effect on our H-R diagrams.}.

\subsection{Discussion of Individual Clusters}
\label{Sec-indiv}

\subsubsection{Ruprecht~44 (C0757$-$284)} 
The Ru~44 cluster has been described by 
Moffat \& FitzGerald (1974), FitzGerald \&
Moffat (1976), Havlen (1976), and
Turner (1981); see also McCarthy \& Miller (1974).
The cluster is a condensation in the Pup~OB2 association.
The WN4.5 star WR10 (HD~65865=MR~11) is listed by Moffat \& FitzGerald
as a member, although it lies well outside the central part of
the cluster. (The ``core" as shown in Fig.~1 of Moffat \& FitzGerald has
a radius of 6.4 arcmin; the W-R star lies 14 arcmin to the SE of its
center.)  Distance estimates have ranged from the large values of
Moffat \& FitzGerald (1974) and FitzGerald \& Moffat (1976) who
proposed distance moduli of 14.1--14.2~mag (6.6--6.8~kpc), to the
smaller value of 13.2~mag (4.3~kpc) found by Havlen (1976). The most
recent and complete study is that of Turner (1981), who
finds a distance modulus of 13.3~mag (4.7~kpc) from main-sequence fitting,
in accord with the H$\beta$ value of Havlen.

Many of the existing spectral types in the cluster are listed by Moffat \&
FitzGerald (1974) and Turner (1981) as uncertain, and we obtained new
spectral information for eight stars.  We find that
the stars described as O-type are generally no earlier than 
B-type; for instance
the star LSS~909 described as ``O6:e" by Moffat \& FitzGerald, and revised
to ``O8" by Turner, is actually a B1~V according to our CCD spectroscopy, and
is likely a foreground object. (It was included as one of the outlying 
possible members by Turner.)  Similarly, the 
star LSS~899 (Ru44-182) was
classified by Moffat \& FitzGerald  (1974) as ``O8~V", but subsequently
reclassified as B0~V by 
Reed \& FitzGerald (1983).  On the other hand, the star LSS~891 (Ru44-183)
was called ``O9.5" by FitzGerald \& Moffat (1976) is actually an O8~III(f)
according to our spectroscopy, similar to the O8~V 
classification by Reed \& FitzGerald (1983).
Our derived distance modulus of 13.35~mag is in
excellent accord with that found by Turner (1981).

\subsubsection{Collinder 228 (C1041$-$597)}
Cr 228 is just south of the $\eta$ Carinae clusters Tr~14
and Tr~16, and Cr~232; see Figure 1 in Massey \& Johnson (1993).  An excellent
spectroscopic study of this region was conducted by Walborn (1973a, 1982).
In constructing our list of members, we began with the {\it UBV} photometry
of Feinstein, Marraco, \& Forte (1976). (The photometry of HD~93146 comes
from Turner \& Moffat 1980.) Spectral types are from
Walborn (1973a, 1982), as well as Levato \& Malaroda (1981), with
a few additional types from Tapia et al.\ (1988) based upon unspecified
sources.  To this we add our own 16 new spectral types, mostly of stars
with previous spectroscopy; the agreement between different sources is
generally very good.  We eliminate stars thought to be non-members
based on spectra or color, as indicated by Tapia et al.
We also ignore the plethora of stars with late-B or early-A dwarf
spectral types in 
determining the distance modulus or constructing the HRD.
The resulting true distance modulus is 12.45~mag (3.1~kpc)
and the average color
excess $E(B-V)=0.37$.  (Our {\it apparent} distance modulus, 13.6~mag, 
is identical
with that found by Tapia et al. but the {\it true} distance modulus is somewhat
greater since they derive a higher average reddening for Cr~228 based upon
calculating $E(B-V)$ from their measured $E(V-K)$ values.)  Our distance
modulus would place the cluster at essentially the same distance as the
rest of the $\eta$~Car complex; see Massey \& Johnson 1993).
A few of the stars have luminosity classes or reddenings inconsistent
with the adopted distance, but the corrections are minor and probably
all of these stars with spectral types are members.  
The W-R member is WR24, of type WN7.

\subsubsection{Trumpler 14/16}

The LBV star $\eta$~Carinae is part of the Tr 14/16 complex, as is
the W-R star HD~93162=WR25 (WN7+abs).  Massey \& Johnson (1993) provide
a modern CCD study of this region, including spectroscopy of many of the
brightest blue stars, concluding that Tr~14 and Tr~16 were at the identical
distance and were coeval.   A comprehensive study of the
fainter members is given by DeGioia-Eastwood et al.\ (2001), who study
several background fields in order to recognize pre-mainsequence objects
in the HRD.

\subsubsection{Pismis~20 (C1511$-$588)}
The only previous photometry of Pismis~20 is the photographic study
by Moffat \& Vogt (1973).
WR67=HD~134877=Pis~20-8 (WN6) 
is nearly 2 arcmin from the central core of the cluster, which is
heavily concentrated in a region roughly 1 arcminute in diameter.
We derive a distance modulus of 12.7~mag (3.5~kpc) for the cluster, and an
average reddening of E(B-V)=1.1, with little scatter.  The reddening of
WR67 given by Conti \& Vacca (1990) is identical with this value,
and we find $M_v=-5.1$, consistent with that
expected for a WN6 star (Table II of Conti \& Vacca 1990).
We conclude WR67 is a member.  Van der Hucht et al.\ (1981) also list WR66
as a possible member, but its location some 46 arcminutes south of the
cluster makes this extremely unlikely.

\subsubsection{C1715$-$387}
The cluster was studied by Havlen \& Moffat (1977), who identified
15 members and 9 non-members from their photoelectric {\it UBV} photometry.
The cluster appears to contain two Wolf-Rayet stars, WR89 
(=LSS 4065=C1715$-$387-1), of type WN7\footnote{Walborn \& Fitzpatrick (2000)
recently classified this star as ``WN8-A".  After N. Walborn kindly
called the matter to our attention, we re-examined our own higher S/N of
this star, and we find we are in accord with the WN7 classification given
by van der Hucht et al.\ (1981). One may compare the Walborn \& Fitzpatrick (2000) spectrum to that of
the spectrum WN8 ``standard" HD~177230 show by Massey \& Conti (1980).
We would argue that 
the WN7 classification of LSS~4065 is slightly preferred over
WN8 given the lack of He~I P~Cygni
(and general weakness of the He~I emission), plus the relative intensity
of N~III $\lambda\lambda 4634,42$ emission relative to He~II $\lambda 4686$.
}
and WR87(=LSS 4064=C1715$-$387-3),
also of type WN7.  Their spectroscopy identified several early-type O stars,
including two early-O type supergiants: star 2 (=LSS~4067), an O4~If, and 6, an O5~If,
both of which we confirm.  They derive a distance modulus of $12.6\pm0.3$~mag
(3.3~kpc) via  main-sequence fitting and $12.0\pm0.3$~mag (2.5~kpc) via
spectroscopic parallax, adopting a compromise $12.3\pm0.3$~mag [sic] (2.9~kpc).  Th\'{e}, Arens,
\& van der Hucht (1982) obtained Walraven-system photoelectric photometry
of these plus additional members, and conclude that the distance modulus
cannot be determined with any certainty from the photometry alone. 

We have obtained spectra for 5 of the stars, all of which are of O-type,
ranging from O4~If to O9.5~III.  Four out of the five stars have nearly
identical reddenings, with $E(B-V)=1.85\pm0.03$; the O9.5~III has
a smaller value, $E(B-V)=1.65$.  The reddening is peculiar towards C1715$-$387,
or else there is a significant zero-point problem in one (or both) of the
colors given by Havlen \& Moffat (1977).  A value of $q_r$=0.83 is suggested
for most of the stars, but a value nearer the canonical 0.72 is indicated
by the less-reddened O9.5~III star.

If we take only the two O dwarfs, we find a distance modulus of 12.0~mag; the
distances of the supergiants are consistent with this, although they would
nominally suggest a distance modulus of 12.2~mag.  
We adopt a distance modulus of
12.0~mag (2.5~kpc), identical to value obtained by Havlen \& Moffat (1977)
using the same data but with less complete spectroscopic data.

Th\'{e} et al.\ (1982) suggested that one of the stars, (their No.~35) is an M-type
supergiant with rather low absolute luminosity ($M_{bol}=-6.3$, 
adjusted for our
closer distance modulus), where the BC was adopted apparently assuming
that the star is M0.  If so, the star cannot be coeval with the rest
of the cluster, given the HR diagrams we derive subsequently.
 We note that there is no spectroscopy of this star, only
photometry, and the star could simply be a foreground dwarf. Or, if the star
is a {\it late}-type 
M supergiant, then its bolometric luminosity might be a couple
of magnitudes more negative.

\subsubsection{Pismis~24 (C1722$-$343)}
\label{Sec-Pis24}

Moffat \& Vogt (1973) obtained photometry of 15 stars, 12 of which
they concluded are members.  We obtained spectroscopy of
11 stars, 4 of which were not included in their list. 
We find that two of the cluster stars are of type O3, one of which is
a supergiant (HDE 319718=Pis~24-1) 
and the other appears to be a giant (what we now
call Pis~24-17); these stars were discussed in Section~\ref{Sec-o3s}.
Most of the rest are found to be
of O-type, although we do reach B dwarfs.  Thus this
is a highly
interesting cluster, and additional photometry and spectroscopy are highly
warranted.  

For HDE 319718 we note that the $V=10.01$ photometry given by
Crampton (1971) is at variance with the $V=10.43$ photometry
of Moffat \& Vogt (1973) by 0.4 mag.  Additionally, there
 are 0.1~mag differences
in the colors.
Lortet et al.\ (1984)
describe this star as ``O4(f)" and state that the radial velocity changed
by 94~km~s$^{-1}$ during a few days, so possibly the light variation was
real.  
However, there is also photoelectric photometry by Neckel (1978).  His
value of $V=10.27$ is probably consistent with Moffat \& Vogt (1973), as
Neckel used a large aperture that would have included some of the
companion stars; see discussion in Lortet et al. (1984).

The cluster is associated with the nebula NGC~6357, a large shell with
``several roundish nebulae" (Lortet et al. 1984).  The brightest of
these is near Pis~24.

In order to provide {\it V} and {\it B-V} photometry for the four additional
stars, we obtained images in {\it B} and {\it V} with the WIYN 3.5-m on
1998 Sept 29.  The field of view was 6.7 arcmin on a side with a scale of
0.2"/pixel.  At airmasses in excess of 3, the image quality was poor, about
2.6". We obtained photometry differentially relative to the stars with
photoelectric photometry by Moffat \& Vogt (1973).  The 1$\sigma$
scatter from 5 stars
was 0.03.  We refer to the four new stars as 16-19; all four are located
near star HDE~319718, with a pair 15" to the east, and a pair 15" to the north.
The pair to the east we label 
16 (NW of pair) and 17 (SE of pair).  The pair to the north we label 
18 (W of pair) and 19 (E of pair).  These stars are clearly visible in 
Fig~4a of Lortet et al.\ (1984); our star Pis~24-17 is the one they
label ``N35B", and classified as O4-5~V, although our better sky-subtracted
spectrum reveals it is of O3~III(f*) type (Sect.~\ref{Sec-o3s}).

If we use the ten stars with certain luminosity classes
to compute the spectroscopic parallax, we
find a distance modulus of $12.03\pm0.14$ (s.d.m.)~mag.
If we use only the six O dwarfs
to derive the distance, we compute $11.99\pm0.05$ (s.d.m.)~mag.  
We adopt a true distance modulus of 12.0~mag (2.5~kpc),
somewhat further than 
the 1.7~kpc suggested by
Neckel (1978) and Lortet et al.\ (1984).  

The cluster contains the Wolf-Rayet star HD~157504 (WR93), of type WC7.
Such late-type WC stars are unknown in the Magellanic Clouds, and are found
primarily inwards of the sun in the Milky Way.  
Conti \& Vacca (1990) describe this star as a ``WCE+abs" and derive a distance
of 1.1~kpc, considerably closer than then 2.5~kpc we find for the cluster.
This includes a very large correction ($A_V=5.9$~mag) for interstellar
extinction.  The star is located  4 arcminutes west of the
central cluster, which is otherwise extended over 5 arcminutes NS, but
1-2 arcminutes EW.  However, it
is not clear what photometry over a larger field would reveal: at this
distance 5 arcmin corresponds to 4~pc, while stellar drifts over 1~Myr
would extend to 10~pc at 10 km~s$^{-1}$. Thus the cluster might well be larger
than indicated. 
We consider 
HD~157504 a likely member of Pis~24.

\subsubsection{Trumpler 27 (C1732$-$334)} 
Th\'{e} \& Stokes (1970) studied the cluster by means of photoelectrically
calibrated photographic {\it UBV} photometry, identifying many highly
reddened early-type stars, and deriving a true distance modulus of 10.2~mag
(1.1 kpc) based upon main-sequence fitting.  Moffat, FitzGerald, \&
Jackson (1977) obtained photoelectric {\it UBV} data and some objective
prism (and a few slit) spectral types; they derive a distance modulus
of $11.7\pm0.2$~mag (2.1~kpc) based upon spectroscopic parallax.  We note here
that the difference between the photographic and photoelectric {\it V}
magnitudes shows a strong magnitude dependence, and that there are 
significant color terms between the photographic and photoelectric
{\it B-V} values, 
which partially explains the difference in the
derived distances.
Baker \& Th\'{e} (1983) used the Walraven photometric system to determine
a distance modulus of 11.2~mag (1.7~kpc), citing the poorly known
absolute magnitude calibration for supergiants to explain the difference
between theirs and Moffat et al's results.

We have new spectral types for twelve OB stars, eight 
of which are in common with
previous slit spectroscopy by Moffat et al. (1977).  Unfortunately, all but
one of these are giants and supergiants, which intrinsically
exhibit a large range in $M_V$.  For instance, a B0~I could easily range
from $M_V=-6$ to $-7$ (Humphreys \& McElroy 1984).    
Our spectroscopy suggests a
distance modulus of $12.0\pm0.3$~mag (2.5~kpc), slightly greater than the
Moffat et al.\ value. The region is ripe for a CCD study and spectroscopy 
that reaches the dwarfs.

The cluster contains two Wolf-Rayet stars,
WR95 (=Tr~27-28) of type WC9, and WR98 (=MR~76=Tr~27-105) of type WC7/WN6.
Conti \& Vacca (1990) infer their distances as 2.8~kpc and 2.4~kpc,
respectively, consistent with the distance we find from spectroscopic 
parallax of the OB stars.  The region also contains two other evolved
stars: Tr28-1 is of type M0~Ia (Imhoff \& Keenan 1976), 
and Tr28-102 is of type G0~Ia (Moffat et al.\
1977).  Dereddening the stars using the intrinsic
colors of FitzGerald (1970),  we find absolute visual magnitudes of $-7.9$
and $-6.9$. The former is consistent with an extreme M-type supergiant;
Imhoff \& Keenan (1976) estimate the star's absolute magnitude
as $-7.2\pm0.5$ from coude spectroscopy of luminosity-sensitive features.
The $M_V$ expected from a ``G0~Ia" star is $-7.5$ (cf. Humphreys \& McElroy 1984), although G supergiants cover a 5~mag range in luminosity depending
upon whether the ``a" could be a ``b" or not!  We conclude that a value
$(m-M)_o=12.0$ is consistent with both the W-Rs and the two other supergiants
as being members.

The data here suggest a significant age spread for the cluster.
The G0~Ia star falls near the 10~Myr isochrone, and
the two B8I stars classified
have ages of 6~Myr, considerably
older than the 2-4~Myr age suggested by the rest of the stars.  

We suggest that this apparent ``age spread" is due to the difficulty
of separating real cluster members from background objects. For instance,
the two early B-type stars Tr27-44 and Tr27-107 both have absolute visual
magnitudes corresponding to {\it giants} rather than {\it supergiants} if
we assume they are indeed cluster members.  
However, their spectra both indicate
an extremely high luminosity class (Ia), 
with a rich assortment of strong,
narrow metal lines.  We believe these two stars are actually background
objects, suggesting that other apparent cluster members may be unassociated
with the cluster instead. We include these two stars in the HRD, as their
implied masses are too low to affect our judgment of the W-R progenitor
masses.

\subsubsection{NGC~6871 (C2004+356)}
The long-period (119 day) Wolf-Rayet binary HD190918 (WN4.5+O9~Ia) is part of
the large NGC~6871 complex.  The region was included in Massey et al.\ (1995a).
No brightness ratio of the O9~Ia to WN4.5 star has been published; however,
even a casual inspection of the spectrum suggests that the O9~Ia star dominates
the continuum.  For the purposes of including the O9Ia star in the HRD, we
make the very conservative assumption that the light of the two components
contributes equally; however, even in the extreme case that we assign {\it all}
of the light to the O9Ia star, this component is not the most massive present
in the cluster.  

We note that the cluster also contains the famous x-ray
binary and black-hole candidate Cyg X-1.  The inferred mass of the O9.7~Ia
component makes it one of the two most massive objects in the cluster,
consistent with the highly evolved state of its companion.  Spectral
analysis of the star led Herrero et al.\ (1995) to conclude that the star
has a mass of 17-20$\cal M_\odot$, considerably less than the $40\cal M_\odot$
we would deduce here, but this is consistent  the fact that other short-period
binaries also show a similar ``mass discrepancy" between the evolutionary
tracks and those inferred from other means (Burkholder, Massey, \& Morrell
1997), suggesting that mass has been lost via Roche-lobe overflow in such
systems.

\subsubsection{Berkeley~86 (C2018+385)}
This cluster was included in the Massey et al.\ (1995a) study, and we have
not acquired any new data.
It contains one of the well known Wolf-Rayet eclipsing binaries, 
V444 Cyg (WN5+O6).  The O6 component is the most massive star in the cluster,
as judged from the HRD, if we assume a ratio of 0.6 for the luminosity of
the WN5 star to that of the O6 star (Cherepashchuk et al.\ 1995).

\subsubsection{Berkeley~87 (C2019+372)}
This cluster was studied photoelectrically and photographically by
Turner \& Forbes (1982).  The WC5 star ST~3 (WR142) is located very near
the cluster center (star 29 in Fig~1 of Turner \& Forbes).  Previously,
spectroscopy had identified a B2~I (Berk87-3) and O9~V (Berk87-25) star
amongst its other members; we have obtained new spectral types for these
and 11 other members.  The implied spectroscopic parallax is 11.0~mag, and
a slightly high $q_r=0.84$ value is found.  All stars fall into a narrow
range in $E(B-V)$.   We find one of our stars is a B[e] star.

\subsubsection{Cyg~OB2}
This association contains many early-type stars, including one
of the rare O3If* (Walborn 1973b). It was scrutinized both photometrically
and spectroscopically by 
Massey \& Thompson (1991).  Van der Hucht et al.\ (1981) list three W-R stars
as members, but of these only WR144 lies near the 50 arcmin (EW) by 40 arcmin
(NS) region examined
by Massey \& Thompson. WR145 lies 30 arcmin south and 16 arcmin west 
of the concentration of
bright blue stars (Fig.~9 of Massey \& Thompson), and WR146 lies 29 arcmin
east.  The {\it V} magnitude of  WR145 is consistent with membership, although
the lack of color information and wide-range of reddenings makes membership
difficult to determine.   
Only WR145 has line-free photometry, and its estimated distance is
0.5~kpc (Conti \& Vacca 1990), considerably at variance with the 1.7~kpc
distance derived by Massey \& Thompson.  For the sake of this study, we will
make the conservative assumption that only WR144 is a likely member.  

Massey \& Thompson (1991) argue that the star VI~Cyg~No.~12 has many of
the characteristics of an LBV: it is extremely luminous bolometrically
($M_{\rm bol}\approx -11$) and its absolute visual magnitude may be unequaled
($M_V\approx -10$).  It is known to be both spectroscopically and photometrically variable.  Humphreys \& Davidson (1994) characterize the star
an ``A-type Hypergiant" (the spectral type is B5~Ie according to Massey 
\& Thompson), and argue that it is ``not [a] full-fledged LBV".  Here we 
continue to consider it at least an ``LBV candidate". 

\subsubsection{Markarian~50 (C2313+602)}
As described by Crampton (1975), the 
Wolf-Rayet star HD~219460 (WN4.5) was found to lie near the center of
a concentration of
early-type stars by Markarian (1951).  Photographic photometry and a
finding chart were
given by Grubissich (1965). Crampton (1975)
obtained spectra of 8 stars (2 of which were foreground), and derived a
distance modulus of 12.0~mag (2.5~kpc).  Turner et al.\ (1983) obtained
photoelectric photometry for some of the stars, and a multitude of
spectra for the W-R star, which was blended with a B-type visual companion
(separation 1 arcsec), which they classify as B1~II.  They find a distance
modulus of $12.75\pm0.12$~mag (3.6~kpc), using a combination of main-sequence fitting
and spectroscopic parallax.
Our new
spectroscopic distance modulus of $12.79\pm0.1$~mag is in good agreement with this value.

\clearpage
\section{The HRDs: Identifying the Highest Mass Stars and Tests of Coevality}

With the reddenings and distances 
of Section~\ref{Sec-dist}, we can now transform to the H-R diagram using
the methods described in Paper~I.  Since we successfully obtained spectral types
for nearly all of the brighter stars in each cluster (e.g., Table~2), we
rely upon the MK type to give us
the effective temperature. Since the reddening
is potentially variable across each cluster, we use the intrinsic color
as a function of spectral type along with the observed photometry to correct
the $V$ magnitude for interstellar extinction, and then use our adopted
distance modulus (Table~1) to determine $M_V$. The bolometric correction
comes from the adopted effective temperature, yielding the bolometric
luminosity.  Reference to the evolutionary tracks of Schaller et al.\ (1992),
computed for $Z=0.020$ provides both the {\it zero-age} masses and the
{\it ages} of the stars.  

We show in Fig.~\ref{fig:hrds} the H-R diagrams for the 8 clusters for which
we have new data; similar diagrams, made with the identical transformations,
can be found for Tr~14, NGC~6871, Berkeley~86, and Cyg~OB2
in Massey et al.\ (1995a). See Massey et al.\ (1995b) and Massey (1998c) for
a discussion of the associated errors.

We can use these data to identify the highest mass stars, and consider whether
the degree of coevality allows the current ``turn-off" masses to have
relevance to the progenitor masses of the associated Wolf-Rayet stars.

We list in Table~3 the highest mass stars in each cluster, along with their
ages, both according to the evolutionary tracks.  As is evident from 
Fig.~\ref{fig:hrds} the highest mass unevolved stars range considerably
from each cluster to cluster, with the youngest cluster (0.3~Myr), Tr~14,
having stars well
in excess of the 120$\cal M_\odot$ highest mass tracks computed by the
Geneva group.  (Using a very
conservative estimate of the mass-luminosity relation,
we estimate that the highest mass star there corresponds to an initial
mass of 185$\cal M_\odot$, with the next highest mass star being 130$\cal M_\odot$.)  The oldest cluster (8~Myr), Markarian 50, contains stars no
more massive than 20$\cal M_\odot$.   This spread in turn-off masses is
considerably larger than we saw in the Magellanic Clouds;  we
discuss this further in Section~\ref{sec-compare}.

We caution the unwary not to use the data in Table~3 to compare the initial
mass functions of these clusters. Although the data discussed in Massey
et al.\ (1995a) are adequate for those purposes, the data presented here
for the 8 clusters with new data are not, primarily due to the scant
amount (and poor quality) of the photometry.  A project is underway at Lowell
Observatory and CTIO to rectify this situation.

The H-R diagrams in Fig.~\ref{fig:hrds} give readers a chance to judge the
extent of coevality of each of these clusters.  In Paper I we also offered
a more rigorous criterion on which to judge the extent of coevality.  We
can use the same test here.  Let us begin by assuming that the typical
{\it error} in assigning a star's location in the H-R diagram corresponds
to {\it one spectral type}. (This error is considerably less than that resulting
from the use of photometry alone for high-mass, hot stars, as shown 
graphically by Massey et al.\ 1995b, Figures 1c and 1d; see also derivation
in Massey 1998c.) We can now ask what fraction of the stars, above some mass,
are consistent with the cluster being strictly coeval (e.g., the data
being consistent with all the high mass stars having been
``born
on a particular Tuesday", as Hillenbrand et al.\ 1993 
put it.)
We again restrict ourselves to those stars of mass 20$\cal M_\odot$ and
above, as there is a systematic difference between the ZAMS and the
transformed locations in the H-R diagram. (See Figure 8 in Paper I and
the corresponding
discussion.)  In addition, let us compare the median age of the three
highest mass stars to the median age of the entire cluster ($>20 M_\odot$).
We give the results in Table~4.

As in Paper~I, we find excellent agreement between our {\it impressions} from
the H-R diagrams, and the {\it quantitative determination} from Table~4.
If the disagreement between the median cluster age for all stars with
masses $>$20$\cal M_\odot$ (as determined for the stars with spectral types),
and the median ages of the three highest mass stars, differs by more than
0.2~dex, we suspect the cluster may not be coeval.   Or, if less than
80\% of the stars are consistent with a maximum age spread of 1~Myr, we
consider the degree of coevality marginal.  Only Trumpler~27 and NGC~6871
fail both of these tests.  As we discussed earlier, this doesn't necessary
reveal that the star-formation process for the massive stars lasted 
significantly longer here than for the other clusters---in the case of Tr~27
we suspect that much of the problem is confusion by background supergiants.
By our stringent criteria we list the coevality of Pismis~20 and Cyg~OB2
as questionable, failing one of the two tests, although inspection of the
HRDs suggest that most of the highest mass stars in these regions are in
fact coeval.

\section{Results: Progenitor Masses, Bolometric Corrections, and Ages}
\label{sec-compare}
\subsection{Progenitor Masses: Does Metallicity Matter?}

In Table~5 we list the progenitor masses of the Milky Way W-R stars in
our sample, adopting the cluster turn-offs from Table~4.  Values for
stars not strictly coeval are included in parentheses.  We do not
include any entries for the two clusters we consider to be non-coeval,
Tr~27 and NGC~6871. We illustrate our results in Fig.~\ref{fig:kathy}, where we include our
data from the LMC and SMC from Paper~I.

We see immediately that the progenitor of W-R stars span a large range
in the Milky Way.  {\it Wolf-Rayet stars in the Milky Way are found in
coeval clusters that have turn-off masses as low as 20$\cal M_\odot$,
and as high as $>120 \cal M_\odot$!}

Previous estimates of the ``minimum mass" to become a W-R star have tended
to be around 40$\cal M_\odot$ (Conti et al. 1983).  The fact that we have
two
clusters, Ruprecht~44 and Markarian~50, both with turn-off
masses of 20--25$\cal M_\odot$, both of which are  highly coeval, and both
of which
contain Wolf-Rayet stars, suggests otherwise.
This result is one we could have anticipated from Paper~I, where we
found that at the lowest metallicity (SMC, $Z=0.005$) the lowest
mass progenitors of W-Rs were $>70\cal M_\odot$.  In the LMC ($Z=0.008$)
progenitor masses ranged as low as 30$\cal M_\odot$.  It is perhaps then
not surprising to find 20--25$\cal M_\odot$ stars in the Milky Way
($Z=0.018$) becoming Wolf-Rayet stars.  

In the LMC we found that the progenitor masses of the early-type WNs
(WNEs) covered a very wide range.  The data for the Milky Way, scant as
they are, suggest much the same (Fig.~\ref{fig:kathy}), 
as Berkeley~86 has a very high turn-off
mass.  On the other hand, its Wolf-Rayet member, V444 Cyg, is a relatively
short-period binary and Roche-lobe induced mass-loss may have affected
its evolution and current spectral type.
It is hard to conclude if the WNE stage has an evolutionary significance
in the Milky Way, other than to say that both of the clusters that contain
the lowest turn-off masses contain WNEs.   

It seems inescapable, though, that the WNL-class, and in particular the
WN7s, are in fact descendants of the highest mass stars. The ``WNL" section of Fig.~\ref{fig:kathy}
includes the 50$\cal M_\odot$ WN6 star MR55; the other four stars are all
of WN7 class. Both the
models and other arguments suggest that {\it some} of
these H-rich Wolf-Rayet stars may actually still be core H-burning objects
(see discussion in Conti et al.\ 1995).  We argue below (Section ~\ref{Sec-below}) that in any event these stars are {\it evolved}; e.g.,
they are not simply very high mass stars with strong stellar winds, as
was found in NGC~3603 and R136 (Massey \& Hunter 1998). 

All three of the WC stars in our sample are found in clusters with
high turn-off masses. While the data are scant, this suggests that the
20$\cal M_\odot$ WNE may not become WCs at Milky Way metallicities.  In
the Magellanic Clouds we did find some WCs in clusters with turn-off masses
of 45$\cal M_\odot$.  There is some overlap in the data (as shown in Fig.~\ref{fig:kathy}), and such a difference would be hard to understand
on the basis of stellar evolution; perhaps we are simply seeing the effects
of small-number statistics.

Finally, both the LBV star $\eta$~Car, and the ``LBV candidate" VI Cyg No.~12
(Massey \& Thompson 1991) are found in clusters with the highest turn-offs.
This was also true for {\it all} of the LBVs, and LBV candidates, in the
Magellanic Clouds (Fig.~\ref{fig:kathy}.  We consider this strong evidence that LBVs are a normal
stage in the evolution of the most massive stars.

\subsection{The BCs}

Let us turn briefly now to the bolometric corrections (BCs).  
Our assumptions here
are the same as in Paper~I, namely that we can place limits on
the bolometric corrections for Wolf-Rayet stars by assuming that
the W-R is at least as luminous (bolometrically) as the highest mass
cluster star, and then comparing this to the absolute visual magnitude
of the W-R, following in the footsteps of Humphreys et al.\ (1985)
and Smith et al.\ (1994).

However, in the Magellanic Clouds, the amount of (bolometric) luminosity
change during the W-R phase was quite modest according to the models, $-1.0$
to $+0.5$ mag, relative to the luminosity at the end of core-H burning.
Here, with the far greater mass-loss rates that characterize the Milky Way,
the evolutionary models predict very large luminosity evolution.  This
is dramatically illustrated in Figs.~5-7 of Schaller et al.\ (1992). At
Milky Way metallicities the models predict a change of 4 magnitudes during
the He-burning phase for stars of 85$\cal M_\odot$! For lower luminosities
the change is more modest, and becomes negligible below 25$\cal M_\odot$.
Nevertheless, we can make some useful comparisons.

In Paper I we found that the BCs of the WNEs
ranged from $\lesssim-4$ for the lowest masses to $-7$ for the highest masses.
Here we have only one WN without a companion confusing the photometry,
and the BC is consistent with what we see for the MC WNEs, $\lesssim4.2$ for
the low-mass HD 65865. 

Crowther, Hillier, \& Smith (1995) analyzed 9 Galactic WNL stars, including
two in our sample.  Model analysis suggests BCs of $-$3.1 for HD~93131,
and $-$3.0 for HD~93162.    For these two stars we find, respectively,
BCs of $\lesssim-4.1$ and $\lesssim-5.4$, with no evolution, and $\lesssim-0.6$ and $\lesssim-1.5$ with
maximum evolution.  We can at least say that there is no conflict with
the ``standard model" calculations, in accord with our findings in Paper~I
that there was good agreement between the models and this empirical method
of finding the BCs. 

\subsection{Ages}
\label{Sec-below}

One of the most glaring facts to emerge from Table~5 is that many of the
evolved stars have ages of $\sim$1~Myr, if they are indeed coeval with their
associated clusters.

Because the mass-luminosity relation is quite shallow for high mass
stars, the H-burning lifetime is not a steep function of mass.
A 120$\cal M_\odot$ star will have a H-burning lifetime of 2.3~Myr
according to the Schaller et al.\ (1992) models; it is not clear if the
1~Myr lifetime is consistent with {\it any} star having evolved past
core-H burning.  What then should we make of these ``evolved" stars? 

Could the ages be wrong?
All four of the youngest regions contain O3 stars,
and as Massey \& Hunter (1998) discuss, there is both a ``hot" and ``cool"
temperature scale for O3 stars, and we have adopted the ``hot" scale
(e.g., Vacca, Garmany, \& Shull 1996).  This will lead to younger ages
than had we adopted the cooler scale, but inspection of Table~3 shows that
this is not a significant factor. For instance, although Tr~14/16 contain
many O3 stars, we would infer the same age of the region were we to use the
multitude of O5-O6 stars present.  However, in Cyg~OB2 we reach
a somewhat different conclusion, as here the O3 star 
does give a slightly
younger
age ($\lesssim1$~Myr) than the other massive stars (1.5~Myr). Since the
O3 class is degenerate, accurate placement of these stars
is not possible without detailed modeling, but in general it appears that
the ages we have derived are reliable.

Alternatively, we need to ask whether or not these ``evolved" objects
are in fact evolved.  In both NGC~3603 and R136 there are peculiar
Wolf-Rayet
stars which  Massey \& Hunter (1998) argue are simply ``super Of" stars---
stars which are still H-burning and relatively unevolved,
but whose extremely high luminosities and
stellar winds result in Wolf-Rayet emission features.  However, the NGC~3603
and R136 W-Rs
are clearly unusual: despite being of ``early" type (WN4.5) they
have H-rich envelopes and very high absolute magnitudes.  However, these
properties are normally what we associate with WN7 stars: H-rich envelopes
and high absolute visual magnitudes.  We remind the reader that the Crowther
(2000) study suggests that different WN spectral subclasses 
result from metallicity.  Perhaps the ``WN7" class in the Milky Way
are extreme examples of Of stars, analogs of the very high mass
objects found in NGC~3603 and R136 (with similar young ages).
One could imagine that such objects, still showing H, have strong
emission spectrum because of high mass-loss rates, and enhanced 
composition at the surface due to mixing from the core.

However, we are dubious of this explanation for one outstanding reason:
$\eta$~Car.  Although described as an ``atypical prototype" of an LBV,
it is very hard to imagine that this peculiar object is still in a normal,
H-burning stage.
The alternative
suggestion that $\eta$ Car is a binary (Damineli, Conti, \& Lopes 1997)
has been refuted by improved data (Davidson et al.\ 2000). 

Let us consider whether or not $\eta$ Car is coeval with its surrounding
cluster.  Although the spectral energy distributions of LBVs are poorly
understood (Humphreys \& Davidson 1994), $\eta$ Car is surrounded by
dust, which has conveniently reprocessed its UV radiation into IR,
making its bolometric luminosity relatively well known: the object
is one of the brightest 20$\micron$ sources in the sky.
Westphal \& Neugebauer (1969) estimate its bolometric luminosity as
$-13.6$; 
Davidson et al.\ (1986) find $-12.3$; we have corrected both values to our
3.1~kpc distance.  We show the upper part of the H-R diagram in
Fig.~\ref{fig:eta}, where we have plotted $\eta$ Car using its more
conservative luminosity, and using the 30,000$^\circ$K effective
temperature adopted by Davidson et al.
This diagram certainly implies that $\eta$ Car is coeval with the
Tr~14/16 cluster.
Furthermore, this also suggests that we are not simply looking at a few
very massive stars that happen to have formed first.
Without evolutionary tracks that extend to higher masses, it is hard
to assign an exact age, but $\eta$ Car's location in the HRD is consistent
with it simply being slightly higher mass than the highest mass O3 star,
and evolved.

\section{Summary}

We have conducted a study of 12 Galactic clusters containing Wolf-Rayet stars
and LBVs, obtaining new spectroscopic data for 8.  Of these, all but two
prove to be highly coeval.  We reach the following conclusions:

\begin{enumerate}
\item Wolf-Rayet stars in the Milky Way are found in clusters containing
a large range of turn-off masses.  The data suggest that at the metallicity
that characterizes the Milky Way  some early-type WN Wolf-Rayet stars come
from progenitors with masses as low as 20--25$\cal M_\odot$.

\item The WNEs may come from a large range in masses, as they do in the
Magellanic Clouds (Paper~I), but this result is uncertain, as the one
high-mass WNE star in our sample is a member of a close binary, V444~Cyg.

\item WN7 stars are found only in clusters with the highest masses.
The youngest of these are only 1~Myr old.  Although these could
still be H-burning objects, ``guilt by association" suggests that these are
in fact evolved massive stars, as the youngest region also contains 
the LBV $\eta$~Car, thought to be an evolved object.

\item $\eta$ Car itself is found to be highly coeval with the rest of
the Tr~14/16 complex, despite the region's young age.  It, and the LBV
candidate VI~Cyg~No.~12, are found in clusters with the highest masses.
This is identical to what we found for the Magellanic Clouds in Paper~I, and
argues strongly that LBVs are a normal stage in the evolution of the most
massive stars.

\item The Galactic WC stars are found in clusters with turn-off masses 
$> 70 \cal M_\odot$. In the Magellanic Clouds we find WCs occurring in clusters
with masses as low as 45~$\cal M_\odot$. We argued in Paper~I that most
WNs thus evolve to WCs.  The data for the Milky Way might suggest that only
the more massive stars become WCs, but the sample size in the Milky Way
is small (3 WCs in 3 clusters) and additional data are needed.

\item The BCs of Galactic W-Rs are hard to determine using the cluster
turn-off methods, as considerable luminosity evolution is expected at the
higher mass-loss rates that characterize the Milky Way luminous stars.
The data are at least consistent with the ``standard model" of Hillier
(1987, 1990) as applied to two of the stars in our sample
(Crowther et al.\ 1995).  In Paper~I we found concluded that there
was excellent agreement, with
the BCs of early WN and WC stars found to be
extreme ($\approx -6$~mag).

\end{enumerate}

We note that much recent work has established the need to extend the
theoretical evolutionary tracks to masses higher than 120$\cal M_\odot$.
Stars with masses estimated to be as high as 160$\cal M_\odot$ have been
found in the R~136 cluster (Massey \& Hunter 1998), and the Galactic clusters
Trumpler~14/16, C1715-387, Pismis~24, and Cyg~OB2 all contain stars whose
luminosities place them above the highest evolutionary track computed
by the Geneva group (120~$\cal M_\odot$).  Future observational work is needed
to extend the H-R diagrams of these and other Galactic clusters, and to 
investigate other coeval regions in the Milky Way and nearby galaxies that
can be used to extend these studies.

\acknowledgements

The majority of the data presented here were obtained during a very pleasant
observing run at Cerro Tololo Inter-American Observatory, and we thank its
excellent support staff.  Additional spectra were obtained during director
discretionary time on Kitt Peak, and we thank
Richard Green for making this time available.  This paper was prepared while
the first author was on sabbatical at Northern Arizona University (NAU), and we
thank Barry Lutz for making an office available and for other
hospitality.  E.W. was supported through
the Research Experiences for Undergraduates program, which was supported by
the National Science Foundation under grant 94-23921 to NAU.  Useful comments
on a draft of the manuscript were provided by Deidre Hunter and Nolan Walborn.

\newpage

\figcaption{Two O3 stars in the cluster Pismis 24.  We classify the upper
spectrum (Pis24-1) as O3~If*, and that of the lower spectrum (Pis24-1B)
as O3~III(f*).
\label{fig:o3s}
}

\figcaption{Early O-type Supergiants in Cl1715-387 (LSS4067 and -6),
and in Cr~228 (HD 93130).
\label{fig:oIs}
}

\figcaption{H-R Diagrams for the 8 clusters with new data.  Solid points
denote stars with spectral types, while open points denote stars with only
photometry. Plus signs indicate particularly uncertain placement. The solid
curves are the $Z=0.020$ evolutionary tracks of  Schaller et al.\ (1992),
with the (initial) masses indicated on the right.  The dashed lines are
isochrones computed from the same models, shown for 2~Myr, 4~Myr,
6~Myr, 8~Myr, and 10~Myr.
\label{fig:hrds}
}

\figcaption{The progenitor masses of evolved stars are shown for the
Milky Way (filled circles), the LMC (open circles), and the SMC (stars).
\label{fig:kathy}
}

\figcaption{The upper most section of the H-R diagram of Trumpler 14.
The five hottest stars are all of type O3, and they are plotted using
the values of Vacca et al.\ (1996); the diagonal lines show where the
stars would lie using the cooler scale of Chlebowski \& Garmany (1991).
The filled circles are stars with spectral types; the open circles are
stars with only photometry, and the location of $\eta$ Car is shown.
The solid lines show the $Z=0.020$ evolutionary tracks of Schaller et
al.\ (1992), and the dashed lines show the isochrones completed for
1~Myr, 2~Myr, and 3~Myr. 
\label{fig:eta} 
}

\newpage
\begin{deluxetable}{l c c c c c c r l l l}
\renewcommand{\arraystretch}{0.6}
\tabletypesize{\footnotesize}
\rotate
\tablewidth{0pc}
\tablenum{1}
\tablecolumns{11}
\tablecaption{Clusters, Derived Distances and Reddenings, Sizes, and W-R/LBV Content} 
\tablehead{
\multicolumn{2}{c}{Cluster}
&  
&  
& \multicolumn{2}{c}{$E(B-V)$}
&  
& Size
& \multicolumn{3}{c}{W-Rs/LBVs} \\ \cline{1-2} \cline{5-6} \cline{9-11}
\colhead{Lynga Designation}
& \colhead{Common Name}
& \colhead{$(m-M)_o$}
& \colhead{$q_r$}
& \colhead{Median}
& \colhead{Range}
& \colhead{Refs.\tablenotemark{a}}
& \colhead{(pc)}
& \colhead{Catalog}
& \colhead{Name}
& \colhead{Type}   
}
\startdata
C0757-284 & Ruprecht 44   &13.4&0.67&0.62&0.5--0.7&New&20 &WR10 & HD65865  &WN4.5 \\
C1041-597 & Collinder 228 &12.5&0.77&0.37&0.2--1.0&New& 25  &WR24 & HD93131  &WN7 \\
C1043-594 & Trumpler 14/16&12.5&0.73&0.53&0.2--0.8&1,2&20  &WR25& HD93162 & WN7+abs \\
          &               &    &    &    &        &   &     & &$\eta$ Car    & LBV    \\
C1511-588 & Pismis 20    &12.7&0.80&1.08&1.0--1.2&New&1&WR67 & MR~55    &WN6 \\
C1715-387 & \nodata      &12.2&0.83&1.85&1.7--1.9&New&4&WR87 & \nodata  &WN7 \\
          &              &    &    &    &        &   & &WR89 & AS~223   &WN7 \\
C1722-343 & Pismis 24    &12.0&0.80&1.73&1.6--1.9&New&4&WR93 & HD157504 &WC7 \\
C1732-334 & Trumpler 27  &12.3&0.76&1.32&1.1--2.5&New&5&WR95 & MR~74    &WC9 \\
          &              &    &    &    &        &   & & WR98 & HDE318016&WN7/C7 \\
C2004+356 & NGC~6871    &11.7&0.71&0.46&0.4--1.1& 3  & 20 &WR133&HD190918&WN4.5+O9.5Ia\\
C2018+385 & Berkeley 86  &11.4&0.72&0.80&0.6--0.9&3& 20 &WR139&V444 Cyg    &WN5+O6  \\
C2019+372 & Berkeley 87  &11.0&0.83&1.62&1.4--1.9&New&7&WR142& ST3    &WC5pec (WO2)\\
\nodata   & Cyg~OB2      &11.2&0.80&1.82&1.2--3.4&4  &25 &WR144&MR110       &WC5     \\
          &              &    &    &    &        &   &   &\nodata &VI Cyg No.12&LBVcand \\
C2313+602 & Markarian 50 &12.8&0.76&0.78&0.7--1.0&New& 5 & WR157& HD 219460-B    & WN4.5   \\
\enddata
\tablenotetext{a}{References for distance and reddenings.  For other references,
see discussion of individual associations in Section~\ref{Sec-indiv}.
(1)--Massey \& Johnson (1993); (2) DeGioia-Eastwood et al.\ 2000; (3)--Massey et al.\ 1995a; (4) Massey \& Thompson (1991).}
\end{deluxetable}
\newpage
\begin{deluxetable}{l l l r r r r l }
\renewcommand{\arraystretch}{0.6}
\tabletypesize{\footnotesize}
\rotate
\tablewidth{0pc}
\tablenum{2}
\tablecolumns{8}
\tablecaption{Catalog of the Brightest Stars in Our Sample}  
\tablehead{
& \multicolumn{2}{c}{J2000.0\tablenotemark{a}}
&
& \multicolumn{3}{c}{Phot.\ from Lit.\tablenotemark{b}} 
& \colhead{Spectral Type and/or} \\ \cline{2-3} \cline{5-7}
\colhead{Star}
&\colhead{$\alpha$}
&\colhead{$\delta$}
&
&\colhead{$V$}
&\colhead{$B-V$}
&\colhead{$U-B$}
&\colhead{Comments\tablenotemark{c}} \\
}
\startdata
\sidehead{Ru~44:} \\
LSS909          & 07 59 22.16\tablenotemark{d} & $-$28 54 23.8\tablenotemark{d} && 10.07 & 0.30 & $-$0.86 & New: B1~V, Nonmember (MF74: O8:e; RF83: B1~V)\\
LSS902=Ru44-185 & 07 58 48.49\tablenotemark{d} & $-$28 23 23.5\tablenotemark{d} && 10.62 & 0.32 & $-$0.66 & RF83: B0 V \\
LSS916=Ru44-187 & 07 59 46.25\tablenotemark{d} & $-$28 44 03.2\tablenotemark{d} && 10.93 & 0.29 & $-$0.60 & WR10=HD65865: WN5 \\
LSS891=Ru44-183 & 07 57 58.55\tablenotemark{d} & $-$28 35 29.4\tablenotemark{d} && 10.93 & 0.29 & $-$0.69 & New: O8 III(f) (FM76: O9.5; RF83: O8 V) \\
LSS885          & 07 57 24.9\tablenotemark{d}  & $-$28 42 07\tablenotemark{d}   && 10.98 & 0.34 & $-$0.62 & RF83: B1 V \\
LSS884          & 07 57 20.79\tablenotemark{d} & $-$28 37 58.3\tablenotemark{d} && 11.16 & 0.28 & $-$0.69 & New: B1~V (RF83: B2Ve) \\ 
LSS897=Ru44-184 & 07 58 42.94                  & $-$28 26 20.3                  && 11.18 & 0.29 & $-$0.66 & RF83: B0 V (MF76: B0 V)\\
LSS907=Ru44-186 & 07 59 08.64                  & $-$28 31 08.0                  && 11.18 & 0.36 & $-$0.61 & New: B0 V (RF83: O9 V) \\
LSS899=Ru44-182 & 07 58 51.84\tablenotemark{d} & $-$28 45 04.2\tablenotemark{d} && 11.30 & 0.27 & $-$0.67 & New: O9 III (MF76: O8; RF83: B0 V) \\
LSS898=Ru44-94  & 07 58 45.79                  & $-$28 32 46.6                  && 11.31 & 0.50 & $-$0.63 & New: Be (FM76: Oe) \\
LSS920          & 08 00 03.26\tablenotemark{d} & $-$28 50 25.7\tablenotemark{d} && 11.38 & 0.24 & $-$0.68 & New: O9.5 V (RF83: O8 V) \\
LSS916SF        & 07 59 49.11                  & $-$28 44 39.6                  && 11.60 & 0.26 & $-$0.68 & south-following companion of WR10 \\
LSS901=Ru44-33  & 07 58 56.89\tablenotemark{d} & $-$28 33 30.2\tablenotemark{d} && 11.63 & 0.37 & $-$0.57 & New: B2 III: (FM76: B0III; RF83: B1 V) \\
LSS908=Ru44-128 & 07 59 12.06                  & $-$28 34 05.0                  && 11.64 & 0.34 & $-$0.62 & New: B0.2 V (MF76: B1 V; RF83: O9~V) \\
Ru44-27         & 07 58 55.53                  & $-$28 35 24.8                  && 11.93 & 0.35 & $-$0.57 & New: B0.5 V \\
LSS906=Ru44-148 & 07 59 05.98                  & $-$28 36 50.9                  && 12.15 & 0.39 & $-$0.55 & New: B1 V (FM76: O9:; RF83: B1 V) \\
LSS903=Ru44-41  & 07 58 58.10                  & $-$28 38 35.9                  && 12.20 & 0.35 & $-$0.52 & RF83: B2 V \\
Ru44-93         & 07 58 45.79                  & $-$28 33 01.2                  && 12.51 & 0.38 & $-$0.35 & MF74: B3: \\
Ru44-19         & 07 58 52.06                  & $-$28 35 06.0                  && 12.55 & 0.34 & $-$0.48 \\
Ru44-2          & 07 58 52.18                  & $-$28 36 04.9                  && 12.63 & 0.39 & $-$0.40 \\
Ru44-3          & 07 58 48.96                  & $-$28 35 47.7                  && 12.65 & 0.41 & $-$0.42 \\
Ru44-24         & 07 58 53.60                  & $-$28 35 03.2                  && 12.79 & 0.37 & $-$0.45 \\
Ru44-59         & 07 58 45.22                  & $-$28 36 51.4                  && 13.18 & 0.38 & $-$0.44 \\
Ru44-112        & 07 58 56.79                  & $-$28 32 50.4                  && 13.56 & 0.40 & $-$0.34 \\
Ru44-102        & 07 58 49.09                  & $-$28 31 28.6                  && 13.57 & 0.41 & $-$0.43 \\
Ru44-114        & 07 58 54.77                  & $-$28 31 29.6                  && 13.72 & 0.38 & $-$0.41  \\
Ru44-60         & 07 58 45.20                  & $-$28 36 47.7                  && 13.72 & 0.40 & $-$0.38 \\
Ru44-40         & 07 59 01.82                  & $-$28 37 52.5                  && 13.81 & 0.39 & $-$0.36 \\
Ru44-14         & 07 58 51.42                  & $-$28 33 30.1                  && 13.83 & 0.44 & $-$0.29 \\
Ru44-113        & 07 58 54.03                  & $-$28 33 08.7                  && 14.06 & 0.38 & $-$0.29 \\
Ru44-51         & 07 58 54.52                  & $-$28 36 14.8                  && 14.31 & 0.36 & $-$0.40 \\
Ru44-6          & 07 58 42.95                  & $-$28 35 02.4                  && 14.80 & 0.52 & $-$0.16 \\
Ru44-88         & 07 58 41.08                  & $-$28 32 43.0                  && 14.82 & 0.49 & $-$0.22 \\
Ru44-84         & 07 58 38.15                  & $-$28 34 22.1                  && 14.97 & 0.64 & $-$0.10 \\
\sidehead{Coll 228:} \\
HD93206=Cr228-33 &10 44 22.91\tablenotemark{d} &$-$59 59 35.9\tablenotemark{d}  && 6.28  & 0.14 &$-$0.80& New: O9.5~I (W73:O9.7Ib:(n); LM81: O9.5Ib:+O9.5III:)\\ 

HD93131=Cr228-3  &10 43 52.26\tablenotemark{d} &$-$60 07 04.0\tablenotemark{d}  &&  6.48 & $-$0.02 & $-$0.88 & WR24 WN7 \\                                                        
HD93130=Cr228-1  &10 44 00.35\tablenotemark{d} & $-$59 52 27.9\tablenotemark{d} &&  8.04 &    0.27 & $-$0.71 & New: O7~II(f) (W73:O6~III(f); LM81: O6~III(f)) \\  
         
HD93222=Cr228-6  & 10 44 36.24\tablenotemark{d} &$-$60 05 29.0 \tablenotemark{d} && 8.08 & 0.08    &$-$0.84   &New: O8~III((f)) (W73: O7~III((f)); LM81: O7 III((f)))\\
            
HD93028=Cr228-27 & 10 43 15.34\tablenotemark{d} &$-$60 12 04.2\tablenotemark{d} && 8.36 & $-$0.06 &$-$0.89  &New: O8.5~III (W73: O9~V; LM81: O8.5~V)\\                    
HD93632=Cr228-92 &10 47 12.49\tablenotemark{d} &$-$60 05 49.8\tablenotemark{d}   && 8.39 & 0.29    &$-$0.73  & New: O5~III(f) (LM81:O5~III(f), W73:O5~III(f)) \\                                    
HD93146=Cr228-65 & 10 44 00.02                 &$-$60 05 11.3                    && 8.41  & 0.00 & $-$0.92  & W73: O6.5~V((f)) (LM81: O6 V) \\                         
HD93191=Cr228-2  &10 44 27.50\tablenotemark{d} &$-$59 53 05.9\tablenotemark{d}   && 8.48  &$-$0.02 &$-$0.18 & LM81: Nonmember (B9.5 V)  \\                                               
HDE305523=Cr228-32 &10 44 29.42\tablenotemark{d}&$-$59 57 18.4\tablenotemark{d}  && 8.49  &0.18    &$-$0.76 &LM81: O8.5 II-III \\                                 
HDE305520=Cr228-4  &10 44 05.83\tablenotemark{d}&$-$59 59 41.7\tablenotemark{d}  &&  8.68 & 0.17 &$-$0.69 &LM81: B1Ib \\                         
HD93027=Cr228-14   &10 43 17.96\tablenotemark{d}&$-$60 08 03.2\tablenotemark{d}  &&  8.72 & 0.00 &$-$0.86 &New: O9~V (W73:O9.5~V; LM81: O9.5~IV) \\                           
Cr228-67 &10 44 00.49 &$-$60 06 01.2          && 8.77  &0.00 &$-$0.82 &New: O9~V (LM81: O9~V)\\                                           
Cr228-88 &10 45 52.00 &$-$60 11 33.2          &&8.79  & 0.14  &0.17     \\                                                          

HDE305438=Cr228-24 &10 42 43.78\tablenotemark{d} &$-$59 54 16.5\tablenotemark{d} && 8.80 &$-$0.01& $-$0.89 &New: O8~V((f)) (LM81: O7.5 V) \\
                 
HDE305536=Cr228-5   &10 44 11.17                  &$-$60 03 21.5                   && 8.94  &0.05 &$-$0.82  &New: O9.5~V (LM81 O8.5 V) \\                                      
HD93056=Cr228-13    &10 43 27.49                 &$-$60 05 54.7                   && 8.97 & $-$0.06& $-$0.78 & LM81: B1Vb:  \\                     
HDE305437=Cr228-23  &10 42 45.18\tablenotemark{d}&$-$59 52 19.68\tablenotemark{d} && 9.06 &   0.02& $-$0.80 & New: B0.5~V (LM81 B0.5 V)   \\                                   
HD93501=Cr228-96    &10 46 22.038\tablenotemark{d}&$-$60 01 18.98\tablenotemark{d} && 9.08 &  0.10& $-$0.67 & LM81: B1.5III: SB2? \\                                             
HDE305524=Cr228-7   &10 44 45.2\tablenotemark{d}  & $-$59 54 41.5\tablenotemark{d}  &&9.28  &0.30 &$-$0.72 & New: O7~V((f)) (LM81: O6~Vn)  \\                                 
Cr228-21            &10 43 57.59 &$-$60 05 28.0&&9.31  &0.02 &$-$0.86 & New: O8.5V (LM81: O7.5Vn)   \\                 
HDE305535=Cr228-25  &10 42 54.68\tablenotemark{d} &$-$59 58 19.7\tablenotemark{d} &&9.39  &0.04 &$-$0.44 & LM81: B2.5 V     \\                                              
HD93647=Cr228-90    &10 47 20.50\tablenotemark{d} &$-$60 12 57.0\tablenotemark{d} &&9.44  &0.11 & 0.15 & LM81: Nonmember (A2: V) \\                                                   
Cr228-12            &10 44 36.88 &$-$59 54 24.9 &&9.47  &0.82 &$-$0.29 & LM81: B2.5 Ia:      \\                                             
HD93576=Cr228-93    &10 46 53.84\tablenotemark{d} & $-$60 04 41.9\tablenotemark{d}&&9.57  &0.25 &$-$0.69 & LM81: O9 V        \\                                               
HDE305534=Cr228-11  &10 44 47.51\tablenotemark{d} &$-$59 57 58.9\tablenotemark{d}&&9.67  &0.13 &$-$0.75 & LM81: B0.5V:+B1V:   \\                                              
HDE305522=Cr228-8   &10 44 19.94\tablenotemark{d} & $-$60 00 05.8\tablenotemark{d}&&9.69  &0.06 &$-$0.76 & LM81: B0.5V:+Comp?   \\                                             
HDE305518=Cr228-22  &10 43 44.00\tablenotemark{d}&$-$59 48 17.9\tablenotemark{d} &&9.71  &0.38 &$-$0.59 & LM81: O9.5IV        \\                                 
HDE305543=Cr228-28  &10 43 10.07\tablenotemark{d}&$-$60 02 11.7 &&9.74  &0.05 &$-$0.77 & New: B1~III (LM81: B1 V+B1 V)     \\                       
HD93097=Cr228-69    &10 43 46.95                 &$-$60 05 50.5 &&9.76 &$-$0.02 &$-$0.81 & LM81: B0V  \\                                                     
Cr228-66            &10 43 59.4:&$-$60 05 14: &&9.79  &0.07 &$-$0.79 & LM81: O9.5V    \\                                                 
HDE305521=Cr228-16  & 10 43 49.50 &$-$59 57 22.4 &&9.81  &0.06 &$-$0.69 & LM81: B0.5V    \\                                                 
HD305519=Cr228-57   & 10 44 11.21 &$-$59 55 30.9 &&9.86  &0.09 & 0.08 & LM81: Nonmember (A2 V)    \\                                                   
HDE305516=Cr228-31  &10 43 15.78                  &$-$59 51 05.3                 &&9.87  &0.06 &$-$0.77 & New: B0.5~V (LM81: B0.5V:)  \\                                      
HDE305539=Cr228-94  &10 46 33.07\tablenotemark{d} &$-$60 04 12.6\tablenotemark{d}&&9.90  &0.27 &$-$0.74 & New: O8.5~V (LM81: O7V; Wal82: O7p)    \\                      
Cr228-39            & 10 44 54.80 &$-$59 56 02.1&&9.92  &0.32 &$-$0.76 & New: O8.5~V((f)) (LM81: O8 V)  \\                                    
HDE305525=Cr228-98  & 10 46 05.70\tablenotemark{d} & $-$59 50 49.3\tablenotemark{d} &&10.00  &0.68 &$-$0.42 & LM81: O6V      \\                                                 
HDE305540=Cr228-91 & 10 47 11.44\tablenotemark{d} &$-$60 11 47.1\tablenotemark{d} &&10.05  &0.09 &$-$0.05 & LM81: Nonmember (A0V)   \\                                                    
Cr228-68           &10 44 00.2: & $-$60 06 10: &&10.16  &0.05 &$-$0.73 & LM81: B1V    \\                                                   
Cr228-35           &10 44 37.39 & $-$60 00 59.6 &&10.18 &$-$0.01 &$-$0.16 & LM81: B9.5V   \\                                                  
HDE305532=Cr228-38 &10 45 34.06 &$-$59 57 26.7\tablenotemark{d} &&10.20  &0.34 &$-$0.74 & W82: O6 V((f)) (LM81)  \\                                
Cr228-29          &10 42 36.44 &$-$60 02 34.5 &&10.21  &0.07 &$-$0.36 & LM81: Nonmember (B9.5 Vp?)  \\                                               
Cr228-36          &10 44 36.99 &$-$60 01 11.4 &&10.23  &0.10 &$-$0.62 & LM81: B0.5: V+B0.5: V:  \\                                           
HDE305528=Cr228-80 &10 45 16.71\tablenotemark{d}&$-$59 54 45.9\tablenotemark{d} &&10.28  &0.13 &$-$0.49 & LM81: B2V             \\                    
HDE305533=Cr228-47 &10 45 13.46 & $-$59 57 54.0 &&10.32  &0.13 &$-$0.51 & LM81: B0.5:Vnn+shell   \\                                  
HDE305515=Cr228-44 &10 43 04.23\tablenotemark{d} &$-$59 51 39.2\tablenotemark{d} &&10.35  &0.09 &$-$0.59 & LM81: B1.5V             \\                                       
Cr228-97           &10 46 22.54 &$-$59 53 20.7 &&10.36  &0.51 &$-$0.64 & W82: O5~V (LM81: O5V)    \\                                   
Cr228-43           &10 43 45.14 &$-$59 53 25.2 &&10.40  &0.22 &$-$0.66 & LM81: B2~V             \\            
Cr228-20           &10 44 15.23 &$-$60 07 53.0 &&10.41  &0.67 &$-$0.22 & T88: B2~V              \\                  
Cr228-89           &10 47 13.28 &$-$60 13 34.3 &&10.43 &$-$0.03 &$-$0.69 & LM81: B2V              \\                                         
Cr228-42\tablenotemark{e} &\nodata &\nodata         &&10.48  &0.66 & 0.20 &                         \\                                        
Cr228-19           &10 44 15.96 &$-$60 09 04.2 &&10.52  &0.09 &$-$0.70 &LM81: B1:V: SB2         \\                                       
HDE305538=Cr228-82 &10 45 46.46\tablenotemark{d} &$-$60 05 13.7\tablenotemark{d}&&10.53  &0.25 &$-$0.53& LM81: B0V              \\                                         
Cr228-87           &10 45 32.42 &$-$60 06 17.6 &&10.55  &0.16 &$-$0.12 &LM81: Nonmember (B9 V)  \\                                
Cr228-40           &10 44 32.94 &$-$59 52 52.8&&10.62  &1.14 & 0.86 &                 \\                                        
Cr228-26           &10 43 14.98 &$-$60 07 47.7&&10.63  &0.21 &$-$0.03& LM81: Nonmember (A0 V)   \\                                                   
HDE305537=Cr228-83 &10 45 44.55 &$-$60 04 23.7 &&10.74  &0.09 &$-$0.60 &LM81: Nonmember (B9.5 V)   \\                                                  
Cr228-46           &10 44 56.71 &$-$60 07 56.3 &&10.74  &0.16 & 0.15 &LM81: Nonmember (A1 V)  \\                                                     
Cr228-30           &10 42 36.22 &$-$59 59 25.8 &&10.80  &0.05 &$-$0.69& LM81: B1.5 V          \\                                           
Cr228-37           &10 45 06.57 &$-$60 00 48.5&&10.81  &0.21 &$-$0.63 &LM81: B2 V         \\                                   
Cr228-81           &10 45 53.58 &$-$60 05 37.2&&10.89 & 0.20 &$-$0.62 &LM81: B0.5 V        \\                                             
Cr228-53           &10 43 51.39 &$-$59 57 20.0 &&10.94 & 0.10 &$-$0.65                     \\                                            
Cr228-95           &10 46 25.47 &$-$60 08 44.5 &&10.98 & 0.02 &$-$0.63& LM81: B0 V          \\                 
Cr228-48           &10 43 48.95 &$-$60 09 00.9 &&11.00 &$-$0.03 &$-$0.62 &LM81: B1.5 V        \\                                            
Cr228-84           &10 45 38.81 &$-$60 04 26.3 &&11.05 &0.25 &$-$0.44                     \\                                            
Cr228-86           &10 45 45.15 &$-$60 06 31.4 &&11.06 &0.98  &1.28                     \\                                            
Cr228-41           &10 44 30.11 &$-$59 52 14.0 &&11.06 &0.21 &$-$0.63                     \\                                            
Cr228-18           &10 44 50.59 &$-$59 55 44.9 &&11.07 &0.25 &$-$0.70                     \\                                            
Cr228-75           &10 43 50.04 &$-$60 01 54.1 &&11.15 &0.32 & 0.19                     \\                                            
Cr228-85           &10 45 34.22 &$-$60 04 31.7 &&11.20 &0.78 & 0.50                     \\                                            
Cr228-49           &10 42 46.22 &$-$60 00 57.5 &&11.20 &0.01 &$-$0.36                     \\                                            
Cr228-78           &10 43 31.62 &$-$60 03 16.2 &&11.44 &0.18 &$-$0.36                     \\                                            
Cr228-77           &10 43 48.87 &$-$60 00 36.8 &&11.57 &0.20 &$-$0.58                     \\                                            
Cr228-74           &10 43 46.88 &$-$60 08 26.4 &&11.66 &0.18 &$-$0.43                     \\                                            
Cr228-61           &10 44 01.0: &$-$59 52 40:  &&11.70 &0.22 &$-$0.59                     \\                                            
Cr228-55           &10 43 39.85 &$-$59 55 16.2 &&11.71 &0.31 & 0.14                    \\                                             
Cr228-51           &10 44 14.54 &$-$60 01 27.3 &&11.88 &0.10 &$-$0.61                     \\                                            
                                
\sidehead{Tr~14/16: See Massey \& Johnson (1993)} \\

\sidehead{Pis 20:} \\
HD134959=Pis20-1 & 15 15 24.07 &$-$59 04 29.2 &&8.20 &0.93 &$-$0.07 &B2.5 Ia \\
Pis20-2          & 15 15 23.82 &$-$59 04 17.9 &&10.45 &0.71: &$-$0.17 &O8.5 I \\
Pis20-3          & 15 15 22.41 &$-$59 04 17.4 &&10.75 &0.90 & $-$0.14 & B0 I \\
Pis20-4          & 15 15 22.41 &$-$59 04 30.3 &&11.37 &0.87 & $-$0.11 & B0.2 III\\
Pis20-5          & 15 15 23.56 &$-$59 03 59.2 &&11.96 &0.89 & $-$0.10 & B0 I-III\\
Pis20-6          & 15 15 19.56 &$-$59 03 24.1 &&11.91 & 1.07 & $-$0.09 & Early B V \\
LSS3327=Pis20-7 & 15 15 17.26  &$-$59 04 48.7 &&11.29 & 0.83 & $-$0.17 & B0 I-III\\
WR67=HD134877=Pis20-8
     & 15 15 32.63  &$-$59 02 30.6 &&11.94 & 0.65 & $-$0.13 & WN6 \\

\sidehead{C1715-387:} \\

LSS4065=C1715-387-1&17 19 00.50 & $-$38 48 52.5 && 11.02 & 1.54 & 0.41 & New: WN7 (HM77: WN8+OB,WF00: WN8-A) \\
LSS4067=C1715-387-2 &17 19l05.51 & $-$38l48 50.5 && 11.16 & 1.54 & 0.37 & New: O4 If+ (HM77: O4f)\\
C1715-387-6         &17 19 05.96 & $-$38 46 46.1 && 11.64 & 1.54 & 0.35 & New: O5 If+  (HM77: O5f)\\
LSS4064=C1715-387-3 & 17 18 52.82 & $-$38 50 04.7&& 12.00 & 1.70 & 0.57 & WN7 (HM77: WN8+OB)\\
C1715-387-8         & 17 19 04.38 & $-$38 49 05.8&& 12.52 & 1.52 & 0.37 & New: O5~V (HM77: O8) \\
C1715-387-12        & 17 18 42.77 & $-$38 49 51.3&& 12.57 & 1.52 & 0.37 & New: O6~If \\
C1715-387-13        & 17 18 47.59 & $-$38 49 58.8&& 12.77 & 1.50 & 0.38 & New: O7 V((f)) (HM77: O8) \\
C1715-387-9         & 17 19 05.70 & $-$38 49 03.2&& 12.99 & 1.59 & 0.41 & \\
C1715-387-16        & 17 18 53.32 & $-$38 51 14.2&& 13.20 & 1.78 & 0.60 & \\
C1715-387-18        & 17 19 00.73 & $-$38 49 24.2&& 13.41 & 1.46 & 0.30 & \\
C1715-387-20        & 17 18 44.85 & $-$38 50 01.3&& 13.45 & 1.35 & 0.17 & New: O9.5 V \\
C1715-387-10        & 17 19 01.0: & $-$38 49 03:&& 13.64 & 1.52 & 0.30 & \\
C1715-387-19        & 17 19 01.63 & $-$38 49 11.5&& 14.15 & 1.58 & 0.39 & \\
C1715-387-23        & 17 19 04.69 & $-$38 49 51.0&& 15.03 & 1.50 & 0.39 & \\
C1715-387-24        & 17 19 05.57 & $-$38 49 27.8&& 15.64 & 1.21 & 0.30 & \\
\sidehead{Pis 24:} \\
HDE319718=Pis24-1&17 24 43.41& $-$34 11 56.5 &&10.43 & 1.45 & 0.40 & New: O3 If* (C71: O7; LTN84: O4(f)) \\
HD 157504=WR93&17 25 08.79& $-$34 11 12.1 &&(11.46) & (1.15:) & \nodata & WC7(+abs?) \\
Pis24-17  &17 24 44.7:& $-$34 12 02:&&11.84 & 1.49 & \nodata & New: O3 III(f*) (LTN84: O4-5~V; see text for ID)\\
Pis24-2   &17 24 43.20& $-$34 12 43.5&&11.95 & 1.41 & 0.32 & New: O5.5 V((f)) \\
Pis24-15  &17 24 28.86& $-$34 14 50.3&&12.32 & 1.27 & 0.14:: & New: O8 V \\
Pis24-13  &17 24 45.68& $-$34 09 39.2&&12.73 & 1.48 & 0.11 & New: O6.5 V((f)) \\
Pis24-3   &17 24 42.21& $-$34 13 21.0&&12.75 & 1.41 & 0.24 & New: O8 V \\
Pis24-8   &17 24 38.81& $-$34 14 58.2&&12.98 & 1.44 & 0.48 \\
Pis24-10  &17 24 35.94& $-$34 13 59.9&&13.02 & 1.40 & 0.40 & New:O9 V\\
Pis24-16  &17 24 44.3:& $-$34 12 00:&&13.02 & 1.60 & \nodata & New: O7.5 V (See text for ID)\\
Pis24-7   &17 24 47.81& $-$34 15 16.5&&13.46 & 1.68 & 0.58  \\
Pis24-12  &17 24 42.22& $-$34 11 41.1&&13.88 & 1.47 & 0.38 & New: B1 V \\
Pis24-4   &17 24 40.39& $-$34 12 05.9&&13.93 & 1.43 & 0.53 & \\
Pis24-18  &17 24 43.2: & $-$34 11 42:&&13.97 & 1.48 &\nodata &New: B0.5 V: (See text for ID)\\
Pis24-9   &17 24 39.29& $-$34 15 26.4&&14.26 & 1.40 & 0.40 \\
Pis24-11  &17 24 34.68& $-$34 13 17.1&&14.53 & 1.57 & 0.30:: \\
Pis24-19  &17 24 43.5:& $-$34 11 41:&&14.43 & 1.39 & \nodata & New: B1 V (See text for ID)\\
\sidehead{Tr 27:} \\
Tr27-1   &17 36 10.07 & $-$33 29 40.5&& 8.79&3.12 &3.32&New: M0 Ia (MFJ77: M0 Ia)\\
Tr27-1a & 17 36 10.1:& $-$33 29 36  &&12.70&1.55 &0.28& \\
LSS4253=Tr27-2  &17 36 10.74 & $-$33 28 48.1&&10.55&1.28 &0.18&         New: B0Ia (MFJ77: O9 Ia) \\
Tr27-3  &17 36 12.94 &$-$33 28 51.6 &&13.16&1.17 &0.19& \\
Tr27-4  &17 36 13.21 &$-$33 30 05.4 &&11.93&0.59 &0.16& \\
Tr27-5  &17 36 10.42 &$-$33 30 02.3 &&12.16&1.23 &0.08 &        New: B2.5Ib \\
Tr27-8  & 17 36 09.69 &$-$33 30 54.7 &&11.88&1.66 &0.56 & \\
Tr27-10 & 17 36 15.55 &$-$33 31 28.8 &&12.12&1.64 &0.76 & \\
Tr27-11 & 17 36 18.92 &$-$33 31 24.3 &&12.34&0.96 &0.03 & \\
Tr27-12 & 17 36 22.12 &$-$33 31 10.9 &&12.05&1.57 &0.48 & \\
LSS4264=Tr27-13 & 17 36 25.20 &$-$33 31 08.5 &&11.78&0.96&-0.01 & \\
LSS4262=Tr27-14 & 17 36 23.31 &$-$33 31 45.4 &&11.12&1.18& 0.15 & MFJ77: B0Ib  \\
LSS4263=Tr27-16 & 17 36 24.31 &$-$33 33 10.0 &&10.74&1.09&-0.01 &     New:B0.5 Ia (MFJ77: O9.5 II:) \\
Tr27-19 & 17 36 32.17 &$-$33 31 51.5 &&12.74&1.03& 0.06 & \\
Tr27-20 & 17 36 28.72 &$-$33 31 32.9 &&13.63&0.81&-0.16 & \\ 
Tr27-21 & 17 36 35.31 &$-$33 30 13.0 &&12.59&0.90&-0.06 & \\
LSS4266=Tr27-23 & 17 36 27.36 &$-$33 29 35.9&&10.11&1.43& 0.37 &  New: B0.7 Ia (MFJ77: B0.5 Ib) \\
Tr27-25 & 17 36 37.60 &$-$33 27 21.8 &&11.42&1.41& 0.34 & \\
Tr27-27 & 17 36 29.91 &$-$33 26 34.2 &&13.31&2.16& 0.74 & New: O8III((f)) \\
Tr27-28=WR95 & 17 36 19.86 &$-$33 26 12.2 &&13.38&1.77& 0.86 &  WC9 (MFJ77: WN5) \\
Tr27-30  & 17 36 05.55 &$-$33 27 50.8 &&13.79&1.48& 0.65 & \\
Tr27-32 & 17 36 35.23 &$-$33 34 32.8 &&12.98&1.07&-0.18 & New: B1.5: V: \\
Tr27-34  &  17 36 45.14 &$-$33 31 55.0&&12.94&1.03& 0.03 & MFJ77:B1: V: \\
Tr27-36 & 17 36 36.11 &$-$33 30 58.6 &&13.18&1.08& 0.16 & \\
Tr27-37 & 17 36 37.12 &$-$33 31 00.9 &&13.99&1.13& 0.30 & \\
Tr27-40 & 17 36 40.12 & $-$33 28 41.7 &&13.88&0.79& 0.14 & \\
Tr27-41 & 17 36 40.56 & $-$33 28 03.3 &&14.05&0.87& 0.18 & \\
Tr27-42 & 17 36 08.19 & $-$33 28 55.5 &&14.22&1.23& 0.16 & New: B3 V \\
Tr27-43 & 17 36 14.43 &$-$33 29 16.8 &&10.48&1.99& 1.06 & New: B8I (MFJ77: B9 Ia) \\
Tr27-44 & 17 36 34.54 & $-$33 30 16.2 &&12.11&0.92&-0.10 & New: Nonmember? B1.5Ia (MFJ77: B1: II::)\\
Tr27-46 & 17 36 12.81 & $-$33 29 18.9 &&8.79&1.60& 0.65 & New: B8I (MFJ77: B9 Ia)\\ 
Tr27-46a &  17 36 13.6: & $-$33 29 09: &&11.54&1.73& 0.77 & \\
Tr27-46b & 17 36 12.8: & $-$33 29 12: &&12.98&1.34& 0.30 & \\
Tr27-46c &  17 36 13.1: & $-$33 29 05: &&13.28&1.45& 0.33 & \\
Tr27-47 & 17 36 14.25 & $-$33 29 36.1 &&14.34&1.86& 0.95 & \\
Tr27-49  &  17 36 17.33 & $-$33 30 02.0 &&14.32&1.54& 0.69 & \\
Tr27-52  &  17 36 21.74 & $-$33 28 34.5 &&14.14&1.24& 0.27 & \\
Tr27-53  &  17 36 27.46 & $-$33 29 15.8 &&14.10&1.64& 0.50 & \\
Tr27-55  &  17 36 31.87 & $-$33 28 36.4 &&14.43&1.06& 0.13 & \\
Tr27-61  & 17 36 29.50 & $-$33 30 46.0 &&13.90&1.09& 0.24 & \\
Tr27-62  &  17 36 26.38 & $-$33 30 36.3 &&14.28&0.82& 0.11 & \\
Tr27-64  &  17 36 23.05 & $-$33 34 37.3 &&14.95&1.36& 0.42 & \\
Tr27-68  &  17 36 10.73 & $-$33 31 52.2 &&14.33&0.95& 0.18 & \\
Tr27-69  &  17 36 03.51 & $-$33 29 54.1 &&14.23&0.96& 0.20 & \\
Tr27-102 &  17 35 56.31 & $-$33 25 56.2 && 8.39&1.94& 1.71 & New: G0 I (MFJ77: G0 Ia) \\
Tr27-103 &  17 35 32.74 & $-$33 27 41.2 &&10.69&0.99&-0.07 & MFJ77: B1 II \\
LSS4271=Tr27-104 &  17 36 39.75 & $-$33 21 16.6 &&10.69&0.79&-0.29 & New:O8.5 III (MFJ77: O9 III) \\
HDE318016=Tr27-105& 17 37 13.72 & $-$33 27 56.1 &&11.85&1.37& 0.49 & WR98 WC7/WN6  \\
LSS4259=Tr27-106 &  17 36 17.64 & $-$33 36 36.3 &&11.43&1.02& 0.07 & MFJ77: B2 III \\ 
LSS4257=Tr27-107 &  17 36 16.63 & $-$33 38 10.2 &&11.46&0.94&-0.16 & New: Nonmember? B0.5 Ia (MFJ77: B0V) \\

\sidehead{NGC 6871: See Massey et al.\ (1995a)} \\

\sidehead{Berk 86: See Massey et al.\ (1995a)} \\

\sidehead{Berk 87:} \\
HDE229059=Berk87-3 & 20:21:15.37 & 37:24:31.3 &&  8.71&1.52& 0.40& 
New: B1 Ia (TF82: B2 Ia) \\
Berk87-4  & 20:21:19.25 & 37:23:24.3&& 10.92&1.26& 0.22&New:B0.2 III:\\
Berk87-5  & 20:21:18.66 & 37:22:31.0 && 14.65&1.35& 0.34\\
Berk87-7  & 20:21:23.14 & 37:20:06.2 && 13.02&1.11& 0.28 \\
Berk87-9  & 20:21:24.88 & 37:22:48.0&& 12.09&1.33& 0.34&New: B0.5 V\\
Berk87-13 & 20:21:31.56 & 37:20:41.2&& 11.32&1.09& 0.12&New: B0.5 III:\\
Berk87-14 & 20:21:33.56 & 37:25:23.0&& 14.12&1.33& 0.82\\
V439Cyg=Berk87-15 & 20:21:33.60 & 37:24:51.6&& 11.84&1.54& 0.37&New: B[e]\\
Berk87-16 & 20:21:33.49 & 37:24:19.4 && 13.39&1.33& 0.55&New: B2 V\\
Berk87-18 & 20:21:35.27 & 37:29:12.2 && 12.84&1.63& 0.66&New: B1 V\\
Berk87-22 & 20:21:36.80 & 37:24:32.9&& 13.96&1.46& 0.55\\
Berk87-24 & 20:21:38.03 & 37:25:17.1&& 11.48&1.34& 0.37&New: B1 Ib\\
Berk87-25 & 20:21:38.67 & 37:25:15.5 && 10.46&1.29& 0.19&New: O8.5III (TF82: O9 V)\\
Berk87-26 & 20:21:39.70 & 37:25:05.4 && 11.83&1.42& 0.38&New: B0.5 I \\
Berk87-27 & 20:21:39.76 & 37:22:39.4&& 13.51&1.35& 0.53\\
ST3=Berk87-29 & 20:21:44.38 & 37:22:30.3&& 12.96&1.43&-0.29& WR142 WC4\\
Berk87-31 & 20:21:45.90 & 37:22:25.7 && 12.32&1.40& 0.38&New: B1 V\\
Berk87-32 & 20:21:47.35 & 37:26:31.8 && 11.57&1.34& 0.31&New: B0.5 III\\
Berk87-34 & 20:21:51.04 & 37:26:05.3 &&13.32&1.53& 0.63\\
Berk87-35 & 20:21:54.38 & 37:23:32.2&& 13.85&1.43& 0.68\\
Berk87-38 & 20:21:59.97& 37:26:23.7&& 12.44&1.58& 0.52&New: B2 III:\\
\sidehead{Cyg OB2: See Massey \& Thompson (1991)} \\
\sidehead{Mark 50:} \\
HD219460-A &23 15 12.5: &60 27 01:  &&10.7:& 0.52 & \nodata &C75: B0III  (phot from C75, corrected for comp)\\
HD219460-B &            &          &&10.9:& 0.52& \nodata &New: WN4.5 (phot from C75, corrected for comp)\\
Ma50-23 &23 15 16.68 &60 26 07.0    &&10.68 &0.46& -0.41 & New: B1 III (C75: B0.5 III) \\
Ma50-31&23 15 11.97 &60 26 46.7    &&11.21 &0.64& -0.27 &New: B0.5 II (C75: B1 III)\\
Ma50-30A &23 15 13.0: &60 26 21:  &&11.90 &0.58& -0.25 &New: B1.5V (C75: B2 V)\\
Ma50-1 &23 14 59.86 & 60 27 15.3    &&12.33 &0.45& -0.35 & New: B1.5V (C75: B1.5~V) \\
Ma50-30 &23 15 14.37 &60 26 18.4   &&12.42 &0.49& -0.36 &B2V New: B2 V (C75: B2 V)\\
Ma50-31A &23 15 13.7: & 60 27 01:  &&12.93 &0.76&  0.03 &New: B3 V \\
Ma50-25  &23 15 16.48 &60 26 21.8   &&13.79 &0.55& -0.06 &New: B3 V \\
Ma50-26  &23 15 13.64 & 60 26 06.4  &&13.90& 0.66&  0.06 &New: B5 V \\
Ma50-24   &23 15 18.07 &60 26 05.9  &&14.04 &0.60&  0.11 &New: B8 V  \\
\enddata
\tablenotetext{a}{Units of right ascension are hours, minutes, and seconds, and units of declination are
degrees, arcminutes, and arcseconds. Coordinates are measured from Digitized
Sky Survey images.}
\tablenotetext{b}{The references for photometry are as follows. 
Ru~44--Turner (1981); 
Cr~228--Feinstein, Marraco, \& Forte (1976);
Pis~20--Moffat \& Vogt (1973);
C1715-387--Havlen \& Moffat (1977);
Pis~24--Moffat \& Vogt (1973);
Tr~27--Moffat, FitzGerald, \& Jackson (1977);
Berk~87--Turner \& Forbes (1982);
Mark~50--Turner et al.\ (1983).
}
\tablenotetext{c}{The reference for spectral types are: 
C71: Crampton (1971);
C75: Crampton (1975);
MF74: Moffat \& FitzGerald (1974);
FM76: FitzGerald \& Moffat (1976);
HM77: Havlen \& Moffat (1977);
LM71: Levato \& Malaroda (1981);
LTN84: Lortet, Testor, \& Niemela (1984);
MFJ77: Moffat, FitzGerald, \& Jackson (1977);
RF83: Reed \& FitzGerald (1983);
T88: quoted in Tapia et al.\ (1988);
TF82: Turner \& Forbes (1982);
W73: Walborn (1973a);
W82: Walborn (1982);
WF00: Walborn \& Fitzpatrick (2000)
}
\tablenotetext{d}{Coordinates from SIMBAD.}
\tablenotetext{e}{Star Cr228-42 appears to be missing from finding chart of Feinstein et al.\ (1976).}
\end{deluxetable}
\newpage
\begin{deluxetable}{l c c c c c l}
\renewcommand{\arraystretch}{0.6}
\tablewidth{0pt}
\tablenum{3}
\tablecolumns{7}
\tablecaption{Derived Parameters for the Highest Mass Unevolved Stars}
\tablehead{
\colhead{Association}
&\colhead{$\log T_{\rm eff}$}
&\colhead{$M_V$}
&\colhead{$M_{\rm bol}$}
&\colhead{Mass}
&\colhead{Age}
&\colhead{Spectral type/Comment} \\
& & & &\colhead{($\cal M_\odot$)}
&\colhead{log Myr}
}
\startdata

\sidehead{Ruprecht 44:}                                                          
LSS891   &4.570   &  -4.4&  -8.0   & 27   &6.34   & O8 III(f) \\ 
LSS898   &4.545:  &  -4.3&  -7.7:  & 24:  &6.52:  & Be     \\
LSS902   &4.500   &  -4.7&  -7.9   & 23   &6.72   & B0 V    \\           
Ru44-920 &4.540   &  -3.7&  -7.1   & 21   &6.21   & O9.5 V  \\         
LSS907   &4.500   &  -4.3&  -7.4   & 20   &6.73   & B0 V    \\ 

\sidehead{Collinder 228:}                                                                 
HD 93206    &4.498   &  -7.5& -10.6   & 88   &6.37   &  O9.5 I\\ 
HD 93632    &4.657   &  -6.1& -10.2   & 76   &6.12   &  O5 III(f)\\               
HD 93130    &4.601   &  -6.3& -10.1   & 68   &6.33   &  O7 II(f)\\
HDE 305525  &4.639   &  -5.7&  -9.8   & 58   &6.21   &  O6 V \\ 
Cr228-97    &4.664   &  -4.8&  -9.0   & 47   &5.61   &  O5 V\\
HD 93146    &4.627   &  -5.1&  -9.1   & 44   &6.15   &  O6.5V((f))\\
HDE 305524  &4.613   &  -5.2&  -9.1   & 42   &6.29   &  O7 V((f)) \\
HD 93222    &4.570   &  -5.7&  -9.3   & 42   &6.47   &  O8 III((f)) \\
HDE 305523  &4.553   &  -5.6&  -9.1   & 38   &6.52   &  O8.5 III \\
HDE 305532  &4.639   &  -4.4&  -8.5   & 38   &5.68   &  O6 V((f))\\
HDE 305438  &4.585   &  -4.6&  -8.4   & 32   &6.34   &  O8 V \\
HD 93028    &4.553   &  -4.9&  -8.4   & 30   &6.54   &  O8.5 III \\
HDE 305518  &4.540   &  -5.0&  -8.3   & 29   &6.59   &  O9.5 V\\
HD 93027    &4.556   &  -4.8&  -8.3   & 29   &6.53   &  O9 V \\
Cr228-39    &4.571   &  -4.6&  -8.2   & 29   &6.42   &  O8.5 V((f)) \\
HDE 305539  &4.571   &  -4.4&  -8.1   & 28   &6.37   &  O8.5 V\\
Cr228-67    &4.556   &  -4.7&  -8.2   & 28   &6.52   &  O9 V\\
HD 93576    &4.556   &  -4.7&  -8.2   & 28   &6.52   &  O9 V\\
Cr228-21    &4.571   &  -4.2&  -7.8   & 27   &6.26   &  O8.5 V\\
HDE 305536  &4.540   &  -4.7&  -8.1   & 26   &6.59   &  O9.5 V\\
Cr228-66    &4.540   &  -3.9&  -7.3   & 22   &6.37   &  O9.5 V\\
Cr228-12    &4.320   &  -6.1&  -8.0   & 20   &6.90   &  B2.5 Ia:\\

\sidehead {Trumpler 14/16:}
HD 93129AB  &4.705   &  -7.5& -12.1   &$>$120   &5.94   & O3 If* \\                               HD 93250    &4.710   &  -6.7& -11.3   &$>$120   &5.76   & O3 V  \\                                   HD 93205    &4.710   &  -6.1& -10.7   &104      &5.46   & O3 V    \\                             HDE 303308  &4.710   &  -5.9& -10.4   & 93      &5.45   & O3 V   \\                                  HD 93128    &4.710   &  -5.4& -10.0   & 75      &5.49   & O3 V  \\  
HD 93160    &4.630   &  -5.9&  -9.9   & 62      &6.26   & O6 III \\
HD 93204    &4.664   &  -5.4&  -9.6   & 59   &5.86   & O5 V \\
-59 2600    &4.639   &  -5.5&  -9.6   & 55   &6.19   & O6 V\\
HDE 303311  &4.664   &  -5.0&  -9.2   & 51   &5.59   & O5 V\\
-59 2641    &4.639   &  -5.2&  -9.3   & 49   &6.09   & O6 V\\
Tr14-257    &4.679   &  -4.3&  -8.7   & 44   &5.63   & O4 I\\
-59 2603    &4.613   &  -5.2&  -9.1   & 43   &6.29   & O7 V\\
-58 2611    &4.639   &  -4.6&  -8.6   & 39   &5.66   & O6 V\\
Tr14-404    &4.600:  &  -5.0&  -8.8:  & 37:  &6.33:  & Phot only \\
-59 2636    &4.585   &  -5.2&  -8.9   & 37   &6.43   & O8 V \\
Tr14-484    &4.613   &  -4.8&  -8.7   & 37   &6.13   & O7 V\\
HD 93343    &4.613   &  -4.7&  -8.6   & 37   &6.11   &O7 V\\
-58 2620    &4.627   &  -4.3&  -8.3   & 35   &5.71   &O6.5 V\\
Tr14-165    &4.585   &  -4.8&  -8.5   & 34   &6.39   &O8 V\\
Tr14-36     &4.639:  &  -4.0&  -8.1:  & 33:  &5.73:  &Phot only\\
Tr14-593    &4.611:  &  -4.3&  -8.2:  & 33:  &5.74:  &Phot only\\
-59 2635    &4.571   &  -5.0&  -8.6   & 33   &6.48   & O8.5 V\\
Tr14-449    &4.626:  &  -4.0&  -7.9:  & 31:  &5.76:  &Phot only\\
Tr14-203    &4.600:  &  -4.1&  -7.9:  & 29:  &5.79:  & \\
Tr14-359    &4.585   &  -4.0&  -7.7   & 27   &5.83   &O8 V\\
Tr14-117    &4.540   &  -4.8&  -8.2   & 27   &6.59   &O9.5 V\\

\sidehead{Pismis 20:}                                                                   
HD 134959   &4.320  &  -7.9&  -9.8   & 50   &6.56   & B2.5Ia  \\
Pis20-6     &(4.629:)  &  -4.2&  (-8.2:)  & (34:)  & ($<1$)  & Early B \\                Pis20-2     &4.537   &  -5.4&  -8.8   & 34   &6.57   & O8.5 I \\                                Pis20-3     &4.460   &  -5.5&  -8.4   & 26   &6.77   & B0 I\\                                   Pis20-7     &4.480   &  -5.0&  -8.0   & 23   &6.76   & B0 III \\                       Pis20-4     &4.470   &  -5.0&  -7.9   & 23   &6.78   & B0.2 III\\
Pis20-5     &4.480   &  -4.5&  -7.5   & 20   &6.79   & B0 III    \\                        
\sidehead{C1715-387:}                                                                   
LSS4067     &4.679   &  -7.0& -11.4   &120   &6.04   & O4 If+  \\         
C1715-387-6 &4.651   &  -6.5& -10.7   & 95   &6.15   & O5 If    \\         
C1715-387-8 &4.664   &  -5.7& -10.0   & 68   &6.01   & O5 V  \\            
C1715-387-12&4.622   &  -5.5&  -9.5   & 50   &6.28   & O6 If\\
C1715-387-13&4.613   &  -5.3&  -9.2   & 44   &6.30   & O7 V((f))  \\      
C1715-387-9 &4.579:  &  -5.3&  -8.9:  & 38:  &6.45:  &Phot only\\
C1715-387-16&4.520:  &  -5.4&  -8.7:  & 32:  &6.61:  &Phot only\\
C1715-387-18&4.581:  &  -4.4&  -8.1:  & 29:  &6.26:  &Phot only\\
C1715-387-19&4.594:  &  -4.1&  -7.8:  & 28:  &5.80:  & Phot only\\
C1715-387-20&4.540   &  -4.1&  -7.4   & 22   &6.47   &O9.5 V:\\

\sidehead{Pismis 24:}                                                                     
HDE 319718  &4.705   &  -7.3& -11.8   &120   &5.86   & O3 If*   \\           
Pis24-17    &4.707   &  -6.0& -10.5   & 98   &5.46   &  O3 III(f*) \\    
Pis24-2     &4.652   &  -5.6&  -9.8   & 60   &6.11   &  O5.5 V((f)) \\       
Pis24-13    &4.627   &  -5.1&  -9.0   & 43   &6.14   &  O6.5 V((f)) \\ 
Pis24-16    &4.600   &  -5.1&  -8.9   & 39   &6.35   &  O7.5 V \\      
Pis24-3     &4.585   &  -4.8&  -8.5   & 33   &6.37   & O8 V\\
Pis24-15    &4.585   &  -4.8&  -8.5   & 33   &6.37   & O8 V\\
Pis24-10    &4.556   &  -4.5&  -8.0   & 27   &6.48   & O9 V\\
                        
\sidehead{Trumpler 27:}                                                                  
Tr27-27     &4.570   &  -6.9& -10.5   & 81   &6.32   & O8 III((f)) \\
Tr27-23     &4.440   &  -7.5& -10.2   & 64   &6.47   & B0.5 I\\     
Tr27-2      &4.460   &  -6.5&  -9.4   & 42   &6.58   & B0 Ia\\           
Tr27-46     &4.050   &  -8.7&  -9.3   & 35   &6.68   & B8 I\\    
Tr27-104    &4.553   &  -5.1&  -8.6   & 32   &6.54   & O8.5 III\\          
Tr27-43     &4.050   &  -8.2&  -8.9   & 29   &6.76   & B8 I\\
Tr27-14     &4.460   &  -5.7&  -8.5   & 27   &6.75   & B0 Ib\\
Tr27-16     &4.440   &  -5.7&  -8.5   & 26   &6.79   & B0.5 Ia\\
Tr27-1a     &4.563:  &  -3.8&  -7.4:  & 23:  &5.89:  & Phot only\\
Tr27-103    &4.420   &  -5.4&  -7.9   & 21   &6.88   & B1 I\\
                   
\sidehead{NGC 6871:}

HD 190864       &4.601   & -5.5    &-9.3    & 45  & 6.37 & O7 III \\
HD 226868       &4.518   & -6.4    &-9.6    & (40) & (6.51) & O9.7 I\\

HD 227018       &4.613   & -4.9    &-8.8    & 38  & 6.20 & O7 V \\
HD 191201       &4.500   & -5.8    &-8.9    & 35 &  6.62 & B0 V \\
HD 190918 comp  &4.498   & -5.4::  &-8.5::  & 29:: & 6.67:: & O9.5 I W-R comp.\\
HD 227634       &4.470   &  -5.3&  -8.2   & 25   &6.75   & B0.2III\\
HD 190919       &4.420   &  -5.8&  -8.3   & 25   &6.81   & B1 Ib\\
BD+35 3955      &4.420   &  -5.7&  -8.3   & 24   &6.83   &B1 Ib\\

\sidehead{Berkeley 86:}
V444 Cyg comp   &4.630:: &  -5.5:: & -9.6:: & 53: & 6.24: & O6 III W-R comp.\\
HD 228841       &4.613   &   -5.4  &  -9.3  & 45  & 6.31  & O7 V \\
HD 193595       &4.585   &   -4.9   & -8.6  & 34  & 6.39  & O8 V \\
HD 228969       &4.540   &   -5.0   & -8.4  & 30  & 6.59  & O9.5 V \\
HD 228943       &4.500   &   -5.3   & -8.4  & 28  & 6.68  & B0 V \\

\sidehead{Berkeley 87:}                                                                  
HDE 229059      &4.420   &  -7.7& -10.3   & 69   &6.46   &  B1 Ia \\ 
Bk87-25         &4.553   &  -5.6&  -9.1   & 39   &6.51   &  O8.5III \\    
Bk87-15         &4.577:  &  -5.0&  -8.6:  & 34:  &6.45:  &   B[e]  \\                 Bk87-4          &4.470   &  -5.0&  -7.9   & 23   &6.78   &  B0.2III\\         

\sidehead{Cyg OB2:}
CygOB2-516  &4.652   &  -7.4& -11.6   &$>$120   &6.29   &  O5.5 V((f))\\
CygOB2-431  &4.651   &  -7.0& -11.2   &$>$120   &6.16   &  O5 If\\
CygOB2-417  &4.683   &  -7.2& -11.6   &$>$120   &6.03   &  O4 III(f)\\
CygOB2-465  &4.637   &  -7.3& -11.3   &$>$120   &6.32   & O5.5I(f)\\
CygOB2-734  &4.651   &  -6.9& -11.1   &$>$120   &6.15   & O5 If \\
CygOB2-457  &4.705   &  -6.3& -10.8   &114      &5.67   & O3If \\
CygOB2-304  &4.270   & -10.6& -12.2   & 92   &6.43   &   B5
I    e\\
CygOB2-771  &4.613   &  -6.7& -10.6   & 90   &6.26   &   O7
V\\
CygOB2-462  &4.616   &  -6.5& -10.4   & 80   &6.26   &   O6.5
III((f))\\
CygOB2-632  &4.498   &  -7.3& -10.4   & 75   &6.40   &   O9.5
I\\
CygOB2-483  &4.651   &  -6.0& -10.1   & 71   &6.15   &   O5
If\\
CygOB2-448  &4.639   &  -5.5&  -9.6   & 54   &6.17   &   O6
V    ((f))\\
CygOB2-217  &4.601   &  -5.8&  -9.6   & 52   &6.37   &   O7
III ((f))\\
CygOB2-555  &4.585   &  -5.8&  -9.5   & 48   &6.42   &   O8
V\\
CygOB2-138  &4.537   &  -6.2&  -9.6   & 48   &6.49   &   O8.5
I\\
CygOB2- 70  &4.556   &  -5.9&  -9.4   & 46   &6.48   &   O9
V\\
CygOB2-390  &4.585   &  -5.6&  -9.3   & 44   &6.42   &   O8
V\\
CygOB2-480  &4.600   &  -5.4&  -9.2   & 44   &6.37   &   O7.5
V\\
CygOB2- 59  &4.571   &  -5.7&  -9.3   & 44   &6.46   &   O8.5
V\\
CygOB2-531  &4.571   &  -5.6&  -9.2   & 42   &6.47   &   O8.5
V\\
CygOB2-317  &4.585   &  -5.5&  -9.2   & 42   &6.43   &   O8
V\\
CygOB2-745  &4.613   &  -5.1&  -9.0   & 41   &6.28   &   O7
V\\
CygOB2-299  &4.600   &  -5.2&  -9.0   & 40   &6.36   &   O7.5
V\\
CygOB2-534  &4.600   &  -5.2&  -9.0   & 40   &6.36   &   O7.5
V\\
CygOB2-  5  &4.639   &  -4.5&  -8.6   & 39   &5.66   &   O6
V((f))\\
CygOB2-601  &4.518   &  -5.8&  -9.0   & 37   &6.58   &   O9.5
III\\
CygOB2-421  &4.540   &  -5.6&  -9.0   & 37   &6.55   &   O9.5
V\\
CygOB2-455  &4.585   &  -5.1&  -8.8   & 36   &6.42   &   O8
V\\
CygOB2-258  &4.585   &  -4.9&  -8.6   & 35   &6.40   &   O8
V\\
CygOB2-485  &4.585   &  -5.0&  -8.7   & 35   &6.40   &   O8
V\\
CygOB2-611  &4.613   &  -4.4&  -8.3   & 34   &5.73   &   O7
V p\\
CygOB2-556  &4.420   &  -6.5&  -9.0   & 33   &6.70   &   B1
Ib\\
CygOB2-339  &4.571   &  -4.9&  -8.5   & 32   &6.47   &   O8.5
V\\
CygOB2-473  &4.571   &  -4.8&  -8.4   & 31   &6.46   &   O8.5
V\\
CygOB2-696  &4.540   &  -5.1&  -8.5   & 31   &6.59   &   O9.5
V\\
CygOB2-376  &4.585   &  -4.6&  -8.3   & 31   &6.32   &   O8
V\\
CygOB2-378  &4.500   &  -5.4&  -8.5   & 29   &6.67   &   B0
V\\
CygOB2-227  &4.556   &  -4.7&  -8.2   & 28   &6.52   &   O9
V\\
CygOB2-507  &4.571   &  -4.4&  -8.0   & 28   &6.36   &   O8.5
V\\\
CygOB2-716  &4.556   &  -4.6&  -8.1   & 27   &6.51   &   O9
V\\
CygOB2-588  &4.500   &  -5.1&  -8.2   & 26   &6.70   &   B0
V\\
CygOB2-736  &4.556   &  -4.1&  -7.6   & 24   &6.33   &   O9
V\\
CygOB2-470  &4.540   &  -4.3&  -7.7   & 24   &6.55   &   O9.5
V\\
CygOB2-425  &4.500   &  -4.7&  -7.9   & 23   &6.72   &   B0
V\\
CygOB2-145  &4.540   &  -4.2&  -7.6   & 23   &6.52   &   O9.5
V\\
CygOB2-426  &4.500   &  -4.3&  -7.5   & 21   &6.73   &   B0
V\\
CygOB2-429  &4.500   &  -4.2&  -7.3   & 20   &6.72   &   B0
V\\

\sidehead{Markarian 50:}                                                                   
HD219460-A   &4.480   &  -4.7&  -7.7   & 21   &6.79   & B0 III (W-R visual comp) \\                         
Ma50-31     &4.450   &  -4.6&  -7.4   & 19   &6.87   &  B0.5 I\\
Ma50-23     &4.372   &  -4.5&  -6.7   & 14   &7.11   &  B1 III \\
Ma50-30A    &4.350   &  -3.6&  -5.7   & 11   &7.23   &  B1.5 V\\                
Ma50-1      &4.350   &  -2.8&  -4.8   &  9   &7.24   &  B1.5 V \\
\enddata
\end{deluxetable}
\begin{deluxetable}{l c c c r c c c l}
\renewcommand{\arraystretch}{0.6}
\tabletypesize{\footnotesize}
\rotate
\tablewidth{0pt}
\tablenum{4}
\tablecolumns{9}
\tablecaption{Coevality and Cluster Ages and Turn-off Masses}
\tablehead{
\colhead{Association}
&\multicolumn{2}{c}{Median Age (log Myr)}
&
& \multicolumn{2}{c}{Coevality}
&
& \multicolumn{2}{l}{Cluster Turn-off Mass} \\ \cline{2-3} \cline{5-6} \cline{8-9}
& \colhead{All $>20\cal M_\odot$}
& \colhead{Three Highest Mass}
&
& \colhead{Percent}
& \colhead{Conclusion}
&
& \colhead{($\cal M_\odot$)}
& \colhead{Comments}
}
\startdata
Ru 44   & 6.53 &   6.34  & &   100   &  Yes      &&         25 & HRD looks coeval\\                  
Cr 228  & 6.37 &   6.33   & & 86     &   Yes      &&        90  & HRD implies some age spead\\

Tr14/16 & 6.14 &   5.94   && 81      &    Yes     &&       $>$120 \\

Pismis 20 & 6.77 & 6.56  &&  67  & Questionable  &&     50 & HRD looks coeval\\

C1715-387 & 6.22 & 6.04 && 100 & Yes && $>$120 & HRD very coeval\\

Pismis 24 & 6.24 & (5.86) && 100 & Very likely && $>$120 \\ 

Trumpler 27 & 6.72 & 6.47 && 70 & No && 80 & Contam by non-cluster stars?\\

NGC~6871 & 6.64 & 6.37 && 75 & No && 45 \\

Berkeley~86 & 6.53 & 6.39 && 100 & Yes && 50 \\

Berkeley~87 & 6.51 & 6.51: && 100 & Yes && 70 & Only 3 stars in sample \\

Cyg OB2 & 6.42 & 6.16 && 94 & Questionable && $>$120 & Many high mass stars of sim ages \\

Markarian 50 & 6.87 & 6.87 && 100 & Yes && 20 & Used $\geq 15\cal M_\odot$ \\

\enddata
\end{deluxetable}
\begin{deluxetable}{l l c c c c c c c}
\renewcommand{\arraystretch}{0.6}
\tabletypesize{\footnotesize}
\rotate
\tablewidth{0pt}
\tablenum{5}
\tablecolumns{9}
\tablecaption{Progenitor Masses, Bolometric Corrections, and Ages}
\tablehead{
\colhead{Star}
&\colhead{Cluster}
&\colhead{Spectral Type}
&\colhead{Progenitor Mass} 
&\colhead{$M_V$}
&\colhead{$M_{\rm bol}$}
&\multicolumn{2}{c}{Bol. Corr.} 
&\colhead{Ages} \\  \cline{7-8}
& & & \colhead{($\cal M_\odot$)}
&
&\colhead{(TAMS)}
&\colhead{No evol.}
&\colhead{Max. Evol.}
&\colhead{(Myr)}
}

\startdata
\sidehead{WNE:} \\
HD 65865 & Ruprecht 44& WN4.5 & 25 &$-$4.2 & $-$8.5 & $<-4.3$ & $<-4.5$ & 2.1\\
MR 120    & Markarian 50 & WN4.5 & 20 &\nodata &\nodata & \nodata &\nodata & 7.4\\
V444 Cyg    & Berkeley 86 & WN5(+O6) & 50 & \nodata &\nodata &\nodata & \nodata & 2.5\\
\sidehead{WNL:} \\
MR55 & Pismis 20 & WN6 & (50)        & $-$5.1 & ($-9.9$) & ($<-4.8$) & ($<-2.4$)& (3.6) \\
HD 93131 & Collinder 228 & WN7 & 90  & $-$6.8 & $-10.8$&  $<-4.0$ & $<-0.1$ & 2.1\\
HD 93162 & Trumpler 14/16 & WN7+abs &$>$120 &$-$5.9 &$<-11.2$ &$<-5.3$&$<-2.3$& 0.9\\
WR 87    & C1715-387 & WN7 & $>$120 &$-$7.6 &$<-11.2$ & $<-3.6$ & $<-0.6$ &1.1\\
AS 223   & C1715-387 & WN7 & $>$120 &$-$6.7 &$<-11.2$ & $<-4.5$ & $<-1.5$ & 1.1\\
\sidehead{WCE:} \\
ST3 & Berkeley 87 & WC5pec (WO2) & 70 & $-$3.8 & $-10.5$ & $<-6.7$ & $<-3.0$ & 3.2\\
MR110 & Cyg OB2 & WC5 & ($>$120) & \nodata & \nodata & \nodata & \nodata & (1.4)\\
\sidehead{WCL:} \\
HD 157504 & Pismis 24 & WC7 & $>$120 & $-$6.4 & $<-11.2$ & $<-4.8$ & $<-1.8$& 0.7\\
\sidehead{LBV:} \\
$\eta$ Car & Trumpler 14/16 & LBV & $>$120& \nodata & \nodata &\nodata & \nodata & 0.9\\
VI Cyg No.12 & Cyg OB2 & LBVCand  & ($>$120) & \nodata & \nodata &\nodata & \nodata & (1.4)\\
\enddata
\end{deluxetable}



\begin{references}

\reference {} Abbott, D. 1982, ApJ, 259, 282
\reference {} Armandroff, T. E., \& Massey, P. 1991, AJ, 102, 927

\reference {} Barb\'a, R. H., Niemela, V. S., Baume, G., \& 
Vazquez, R. A. 1995, ApJ, 446, 23
\reference {} Baume, G., Vazquez, R. A., \& Feinstein, A. 1999, A\&AS, 137, 233
\reference {} Bohannan, B., \& Crowther, P. 1999, ApJ, 511, 374
\reference {} Bohannan, B., \& Walborn, N. R. 1989, PASP, 101, 520
\reference {} Burkholder, V., Massey, P., \& Morrell, N. 1997, ApJ, 490, 328
\reference {} Cherepashchuk, A. M., Koenigsberger, G., Marchenko, S. V., 
\& Moffat, A. F. J. 1995, A\&A, 293, 142
\reference {} Chlebowski, T., \& Garmany, C.\ D.\ 1991, ApJ, 368, 241
\reference {} Conti, P. S. 1976, Mem.\ Soc.\ Roy.\ Sci.\ Liege, 6$^e$ Ser.~9, 193
\reference {} Conti, P. S. 1982, in Wolf-Rayet Stars: Observations, Physics,
Evolution, ed. C. W. H. de Loore and A. J. Willis (Dordrecht, Reidel), 3.
\reference {} Conti, P. S., Garmany, C. D., deLoore, C., \& Vanbeveren, D.
1983, ApJ, 274, 302
\reference {} Conti, P. S., Hanson, M. M., Morris, P. W., 
Willis, A. J., \& Fossey, S. L. 1995, ApJ, 445, L35
\reference {} Conti, P. S., \& Massey, P. 1989, 337, 251
\reference {} Conti, P. S., \& Vacca, W. D. 1990, AJ, 100, 431
\reference {} Crampton, D. 1971, AJ, 76, 260
\reference {} Crampton, D. 1975, PASP, 87, 523
\reference {} Crowther, P. A. 2000, A\&A, 356, 191
\reference {} Crowther, P. A., \& Dessart, L. 1998, MNRAS, 296, 622
\reference {} Crowther, P. A., Hillier, D. J., \& Smith, L. J. 1995, A\&A, 
293, 403
\reference {} Damineli, A., Conti, P. S., \& Lopes, D. F. 19996, NewA, 2, 107
\reference {} Davidson, K., Dufour, R. J., Walborn, N. R., \& Gull, T. R. 1986,
ApJ, 305, 867
\reference {} Davidson, K., Ishibashi, K., Gull, T. R., Humphreys, R. M.,
\& Smith, N. 2000, ApJ, 530, L107
\reference {} DeGioia-Eastwood, K., Throop, H., Walker, G., \& Cudworth, K. M.
2001, ApJ, in press
\reference {} de Koter, A., Heap, S. R., \& Hubeny, I. 1997, ApJ, 477, 792
\reference {} Drissen, L, Moffat, A. F. J., Walborn, N. R., \& Shara, M. M. 1995, AJ, 110, 2235
\reference {} Esteban, D., \& Peimbert, M. 1995, Revista Mexicana de Astronomia y Astrofisica, Serie de Conferencias 3, 133
\reference {} Feinstein, A., Marraco, H. G., \& Forte, J. C. 1976, A\&AS, 24, 389
\reference {} FitzGerald, M. P, \& Moffat, A. F. J. 1976, A\&A, 50, 149
\reference {} Grubissich, Cl. 1965, Zs f Ap 60, 256
\reference {} Havlen, R. J. 1976, A\&A, 48, 193
\reference {} Havlen, R. J., \& Moffat, A. F. J. 1977, A\&A, 58. 351
\reference {} Herrero, A., Kudritzki, R. P., Gabler, R.,
Vilchez, J. M, \& Gabler, A. 1995, A\&A, 297, 556
\reference {} Hillenbrand, L. A., Massey, P., Strom, S. E., \& Merrill, K. M.
1993, AJ, 106, 1906
\reference {} Hillier, D. J. 1987, ApJS, 63, 947
\reference {} Hillier, D. J. 1990, A\&A, 231, 116
\reference {} Humphreys, R. M., \& Davidson, K. 1994, PASP, 106, 1025
\reference {} Humphreys, R. M., \& McElory, D. B. 1984, ApJ, 284, 565
\reference {} Humphreys, R. M., Nichols, M., \& Massey, P. 1985, AJ, 90, 101
\reference {} Imhoff, C. L., \& Keenan, P. C. 1976, ApJ, 205, 455
\reference {} Kudritzki, R. P. 1980, A\&A, 85, 174
\reference {} Levato, H, \& Malaroda, S. 1981, PASP, 93, 714
\reference {} Lortet, M. C., Testor, G., \& Niemela, V. 1984, A\&A, 140,
24
\reference {} Lundstrom, I., \& Stenholm, B. 1984, A\&AS, 58, 163
\reference {} Maeder, A. 1997, A\&A, 321, 134
\reference {} Maeder, A. 1999, in Wolf-Rayet Phenomena in Massive Stars and
Starburst Galaxies (San Francisco: ASP), 177
\reference {} Maeder, A., \& Conti, P. S. 1994, ARA\&A, 32, 227
\reference {} Markarian, B. E. 1951, Contr. Burakan Obs. 9, 1
\reference {} Massey, P. 1996, in Wolf-Rayet Stars in the Framework of
Stellar Evolution, ed. J. M. Vreux, A. Detal, D. Frainpoint-Caro, D. Gosset, \&
G. Rauw (Liege: Institut d'Astrophysicque), 361
\reference {} Massey, P. 1998a, ApJ, 501, 153
\reference {} Massey, P. 1998b, in Stellar Astrophysics for the Local
Group, ed. A. Aparicio, A. Herrero, \& A. Sanchez (Cambridge, Cambridge
University Press), 95
\reference {} Massey, P. 1998c, in The Stellar Initial Mass Function,
ed. G. Gilmore \& D. Howell (San Francisco, Astronomical Society of the Pacific), 17
\reference {} Massey, P., Bianchi, L., Hutchings, J. B., \& Stecher, T. P. 1996,
ApJ 469, 629
\reference {} Massey, P., \& Conti, P. S. 1980, ApJ, 242, 638
\reference {} Massey, P., \& Hunter, D. A. 1998, ApJ, 493, 180
\reference {} Massey, P., \& Johnson, J. 1993, AJ, 105, 980
\reference {} Massey, P., Johnson, K. E., \& DeGioia-Eastwood, K. 1995a,
ApJ, 454, 151
\reference {} Massey, P., \& Johnson, O. 1998, ApJ, 505, 793
ApJ, 454, 151
\reference {} Massey, P., Lang, C. C., DeGioia-Eastwood, K., and
Garmany, C. D. 1995b, ApJ, 438, 188
\reference {} Massey, P., \& Thompson, A. B. 1991, AJ, 101, 1408
\reference {} Massey, P., Waterhouse, E., \& DeGioia-Eastwood, K. 2000, AJ,
119, 2214
\reference {} McCarthy, C., \& Miller, E. 1974, AJ, 79, 1296
\reference {} Meynet, G., Maeder, A., Schaller, G., Schaerer, D., \& Charbonnel,
C. 1994, A\&AS, 103, 97
\reference {} Moffat, A. F. J. 1982, in Wolf Rayet Stars:
Observations, Physics, Evolution, ed. C. W. H. de Loore and A. J. Willis 
(Dordrecht: Reidel), 515
\reference {} Moffat, A. F. J., \& FitzGerald, M. P. 1974, A\&A, 34, 291
\reference {} Moffat, A. F. J., FitzGerald, M. P., \& Jackson, P. D. 1977,
ApJ, 215, 106
\reference {} Moffat, A. F. J., \& Vogt, N. 1973, A\&AS, 10, 135
195
\reference {} Neckel, Th. 1978, A\&A, 69, 51
\reference {} Perry, C. L., Hill, G., \& Christodoulou, D. M. 1991, A\&AS, 90,
\reference {} Reed, B. C., \& FitzGerald, M. P. 1983, MNRAS, 205, 241
\reference {} Russell, S. C., \& Dopita, M. A. 1990, ApJS, 74, 93
\reference {} Schaller, G., Schaerer, D., Meynet, G., \& Maeder 1992, A\&AS, 96, 269
\reference {} Schild, H., \& Maeder, A. 1984, A\&A, 136, 237
\reference {} Simon, K. P., Kudritzki, R. P, Jonas, G., \& Rahe, J. 1983, A\&A,
125, 3
\reference {} Smith, L. F. 1968, MNRAS, 141, 317
\reference {} Smith, L. F., Meynet, G., \& Mermilliod, J.-C. 1994, A\&A, 287, 835
\reference {} Sung, H., Bessell, M. S., \& Lee, S-W 1998, AJ, 15, 734
\reference {} Tapia, M, Roth, M., Marraco, H., \& Ruiz, M. T. 1988,
MNRAS, 232, 661
\reference {} Th\'{e}, P. S., Arens, M., \& van der Hucht, K. A. 1982,
Astrophy. Let. 22, 109
\reference {} Th\'{e}, P. S., \& Stokes, N. 1970, A\&A, 5, 298
\reference {} Turner, D. G. 1981, AJ, 86, 222
\reference {} Turner, D. G., \& Forbes, D. 1982, PASP, 94, 789
\reference {} Turner \& Moffat 1980, MNRAS, 192, 283
\reference {} Turner, D. G., Moffat, A. F. J., Lamontagne, R., \& Maitzen, H. M. 1983, AJ, 88, 1199
\reference {} Vacca, W. D., Garmany, C. D., \& Shull, J. M. 1996, ApJ, 460, 914
\reference {} van der Hucht, K. A., Conti, P. S., Lundstrom, I., \& Stenholm, B.
1981, Space Science Rev. 28, 227
\reference {} Vazquez, R. A., \& Feinstein, A. 1990, Rev.\ Mex.\ Astron.\ Astrof., 21, 346
\reference {} Vijapurkar \& Drilling 1993, ApJS 89, 293
\reference {} Walborn, N. R. 1971, ApJ, 167, L31
\reference {} Walborn, N. R. 1973a, ApJ, 179, 517
\reference {} Walborn, N. R. 1973b, ApJ, 180, L35
\reference {} Walborn, N. R. 1982, AJ, 87, 1300
\reference {} Walborn, N. R. 1994, in The MK Process at 50 Years,
ed. C. J. Corbally, R. O. Gray, and R. F. Garrison (San Francisco: ASP), 84
\reference {} Walborn, N. R., \& Fitzpatrick, E. L. 1990, PASP, 102, 379
\reference {} Walborn, N. R., \& Fitzpatrick, E. L. 2000, PASP, 112, 50
\reference {} Westphal, J. A., \& Neugebauer, G. 1969, ApJ, 156, L45
\reference {} Willis, A. J., Schild, H., \& Smith, L. J. 1992, A\&A, 261, 419
\end{references}
\end{document}